\documentclass[11pt]{article}

\usepackage[preprint]{acl}

\usepackage{times}
\usepackage{latexsym}

\usepackage[T1]{fontenc}

\usepackage[utf8]{inputenc}

\usepackage{microtype}

\usepackage{inconsolata}

\usepackage{graphicx}

\usepackage{amsmath}
\usepackage{amssymb}
\usepackage{amsthm}
\usepackage{mathtools} 
\usepackage{bbm}
\usepackage{dsfont}
\usepackage{multirow}
\usepackage{enumitem}
\usepackage{algorithm}
\usepackage{algorithmicx}
\usepackage{algpseudocode}
\usepackage[most]{tcolorbox}
\usepackage{pifont}
\usepackage{tabularx}
\usepackage{booktabs}

\newtheorem{definition}{Definition}
\newtheorem{assumption}{Assumption}

\definecolor{sysblue}{RGB}{30, 58, 95}
\definecolor{sysbg}{RGB}{235, 242, 252}
\definecolor{userteal}{RGB}{20, 78, 68}
\definecolor{userbg}{RGB}{232, 248, 243}
\definecolor{codegray}{RGB}{245, 245, 245}
\definecolor{tagcolor}{RGB}{140, 30, 30}
\definecolor{placeholdercolor}{RGB}{30, 90, 170}

\tcbset{
  promptbase/.style={
    breakable,
    enhanced,
    fontupper=\ttfamily\scriptsize,
    boxrule=0.4pt,
    left=6pt, right=6pt, top=4pt, bottom=4pt,
    before upper={\setlength{\parindent}{0pt}\setlength{\parskip}{2pt}},
  }
}

\newtcolorbox{systemprompt}[1][System Prompt]{
  promptbase,
  colframe=sysblue, colback=sysbg,
  coltitle=white, colbacktitle=sysblue,
  title={\scriptsize\bfseries\sffamily #1},
  attach boxed title to top left={yshift=-0.25mm, xshift=2mm},
  boxed title style={colframe=sysblue, colback=sysblue,
    sharp corners, boxrule=0pt, left=2pt, right=2pt, top=0pt, bottom=0pt},
}

\newtcolorbox{userprompt}[1][User Prompt]{
  promptbase,
  colframe=userteal, colback=userbg,
  coltitle=white, colbacktitle=userteal,
  title={\scriptsize\bfseries\sffamily #1},
  attach boxed title to top left={yshift=-0.25mm, xshift=2mm},
  boxed title style={colframe=userteal, colback=userteal,
    sharp corners, boxrule=0pt, left=2pt, right=2pt, top=0pt, bottom=0pt},
}

\newcommand{\placeholder}[1]{%
  {\ttfamily\color{placeholdercolor}\{#1\}}}

\definecolor{icodebg}{gray}{0.93}
\newcommand{\icode}[1]{{%
  \setlength{\fboxsep}{1.5pt}%
  \colorbox{icodebg}{\ttfamily\small #1}%
}}

\title{Offline A/B Testing of Slate Recommendation Systems with LLMs: Reducing the Dependency on Pre-Collected User Interaction Data}

\author{
 \textbf{Baptiste Bonin\textsuperscript{1,2}},
 \textbf{Maxime Heuillet\textsuperscript{1,2}},
 \textbf{Audrey Durand\textsuperscript{1,2,3}},
\\
 \textsuperscript{1}Mila (Quebec AI Institute),
 \textsuperscript{2}Université Laval,
 \textsuperscript{3}CIFAR AI Chair
\\
 \small{
   \textbf{Correspondence:} \href{mailto:baptiste.bonin.1@ulaval.ca}{baptiste.bonin.1@ulaval.ca}
 }
}

\begin{document}
\maketitle

\begin{abstract}
    Slate recommender systems (RecSys) present users with ordered sets of interacting items (e.g., playlists). We investigate whether large language models (LLMs) can articulate pairwise preferences between slates for synthetic A/B testing of slate RecSys. We introduce a validation protocol measuring the alignment of synthetic preferences with classical RecSys metrics and their compliance with preference axioms, and use it to characterise how LLM pre-training and configuration affect slate preference articulation. Combined with the generalized Rao-Kupper model, synthetic LLM-based A/B testing recovers rankings that remain stable across utility weightings, whereas off-policy estimators are reliable only when the target utility matches the logged behavior. We position it as a screening stage between off-policy evaluation and live experiments: not a replacement for A/B testing, but a way to reserve its cost for the most promising candidates.
\end{abstract}

\section{Introduction}


Slate recommender systems (RecSys) jointly select and order the sets of items (referred to as \textit{slates}) presented to users, accounting for inter-item interactions that define the overall relevance of the list.  
In this paradigm prevalent on platforms such as YouTube \citep{ieSLATEQTractableDecomposition2019a}, Amazon \citep{DingWhole2019}, and Spotify \citep{chenRecsysChallenge20182018} user feedback is typically elicited for the entire slate rather than for individual items. 

Live A/B testing remains the gold standard for comparing multiple RecSys candidates~\citep{Castells_Moffat_2022}: 
users are presented with pairs of recommendations produced by different RecSys and their articulated preferences are used to extract a ranking of the candidates. Unfortunately, the size of the required user pool and the time scale are combinatorial with the number of RecSys candidates to compare, resulting in a costly deployment~\citep{GilotteOffline2018}, motivating offline alternatives. 
Off-policy estimators re-weight logged interactions to estimate candidate performance without deployment. Inverse Propensity Scoring (IPS)~\citep{RosenbaumIPS1983, liUnbiasedOfflineEvaluation2011} suffers from high variance under poor coverage of the combinatorial slate space~\citep{Swaminathan2016OffpolicyEF}; slate-specific refinements~\citep{mcinerneyCounterfactual2020, chaudhariDistributional2024} reduce this variance but, like IPS, require logged exposure data. 
User-behavior simulators~\citep{rohde2018recogym, shiVirtualTaobaoVirtualizingRealWorld2019, zhaoKuaiSimComprehensiveSimulator2023} 
also fit click and rating dynamics from logged interactions, ignoring semantic content of items and slates (titles, description, meta-data). Large Language Models (LLMs) have recently been explored to simulate single-item ratings~\citep{coreccoSUBERRLEnvironment2024}, as they can reason directly from item semantics (e.g., titles, descriptions) without requiring previous user interactions with those specific items. However, LLMs are known to exhibit difficulties in such grading tasks~\citep{zhengJudgingLLMasaJudgeMTBench2023}. 

However, LLMs have shown much stronger alignment with human judgments in pairwise preference tasks~\citep{ tripathiPairwisePointwiseEvaluating2025}. 
We therefore ask:
\textit{Can LLMs enable the offline evaluation of slate recommendation systems by simulating user preferences in synthetic A/B testing?} 

\textbf{Contributions:} 
We introduce a protocol to assess the coherence of preferences articulated by synthetic evaluators in slate A/B testing by measuring their alignment with classical RecSys metrics along with their compliance with preference axioms. Our results provide intuition on LLM pre-training and configuration affinities with slate preference articulation. Finally, we conduct RecSys-ranking experiments showing that synthetic LLM-based A/B testing recovers rankings that remain stable across utility weightings, whereas off-policy estimation is reliable only when the target reward matches the logged behavior; unlike off-policy estimators, it requires no logged exposure or propensity data for the candidates under evaluation, although pre-collected interactions remain necessary to build user contexts and to validate the judge. We do not propose to replace live A/B testing: it is a screening stage that reserves the cost of an online experiment for candidates surviving offline evaluation.


\section{Related Work}
\label{section:related-work}

LLMs are known to align well with human preferences in pairwise comparisons of text responses~\citep{zhengJudgingLLMasaJudgeMTBench2023,chen2024chateval,qinLargeLanguageModels2024, tan2025can,tripathiPairwisePointwiseEvaluating2025}. However, it is unclear whether these results generalize to slate comparisons, as articulating preferences on slates requires to assess their structure (ordering, diversity, progression) and inter-items interactions. 

User preference on slates depends on complementary dimensions at the slate-level that go beyond item-level prediction accuracy. Several metrics have been proposed to capture these dimensions, including serendipity~\cite{Degemmis2015An}, novelty~\cite{Zhang2013The}, and diversity \cite{Zhang2008Avoiding}. Recent work shows that LLMs perform strongly when aligned with these objectives individually, consistently matching or outperforming conventional baselines to evaluate serendipity \cite{kang2024exploring,tokutake2024universal} and  novelty \cite{microsoft2024optimizing,lin2024evaluating}. Our study empirically evaluates the simultaneous alignment with a variety of RecSys metrics to capture the multiple objectives required to provide a realistic proxy for A/B testing.

Beyond preference-articulation, LLMs have been considered as part of the recommender pipeline itself, for example, to increase diversity in re-ranking tasks~\cite{carraro2024enhancing,gao2024llm4rerank}. \textit{LLM for recommendation} in a different line of work from \textit{LLM for RecSys evaluation}, which is the focus of this work. The work closest to ours is probably \citet{wu2024recsys}, which also investigates the potential of LLMs for ranking item RecSys candidates (instead of slate RecSys). We leverage LLMs to simulate user preferences in synthetic A/B testing, using LLM-articulated preferences (in place of user-articulated preferences) within well-defined and commonly used statistical models that enjoy statistical confidence intervals. On the other hand, \citet{wu2024recsys} replace the complete RecSys ranking pipeline by an LLM, losing statistical interpretability in the process.

\section{Problem setting}
\label{section:problem_setting}

We first describe the general problem of slate recommendation as background. We then present the considered problem of ranking slate RecSys candidates and formulate the proposed offline (synthetic) A/B testing evaluation process.

\subsection{Slate recommendation}

A classical recommendation setting is defined by a set $\mathcal I$ of item features (e.g., title, category/genre, optional short description) and a set $\mathcal X$ of user.
To alleviate the text, we often refer to items and users as their features, e.g., ``user $x\in\mathcal X$'', ``item $I\in\mathcal I$''.
A slate is an ordered sequence of $K$ distinct items drawn from the set $\mathcal I$. The set of possible slates is noted $\mathcal L$. In a slate recommendation problem, a slate recommender system (RecSys) sequentially observes a user $x \in \mathcal X$, recommends a slate $L \in \mathcal L$ (given $x$), and observes feedback $R$, either directly provided on the slate or derived from per-item interactions $\mathcal R$ (e.g., ratings, clicks, plays). The objective of a RecSys is to maximize user satisfaction.

It is common to assume that each user $x$ is characterized by its latent \textit{utility function} $u_x : \mathcal L \to \mathbb R$, which assigns a numeric value to every slate \cite{kaufmann2025responserank, xu2025strong}. This utility function is typically unknown and cannot be properly described even by the user. We can only observe signals as a proxy to the utility function. Signals can correspond to feedback on recommendations. 
Alternatively, signals can materialize as preferences obtained from pairwise comparisons between two slates:
\begin{definition}[Preference]
\label{def:pref}
    The preference $\succ_x$ is induced by the utility function such that $L \succ_x L'$, $L \sim_x L'$ (tie), or $L \prec_x L'$ according to whether $u_x(L) - u_x(L')$ is positive, zero, or negative.
\end{definition}
We assume that preferences follow well-established properties
~\citep{rosenDiscreteMathematicsApplications}: 
\begin{assumption}[Preference axioms]
\label{ass:pref_props}
Preferences respect irreflexivity (i.e., a slate is never be preferred to itself $L \not\succ L$), asymmetry (i.e., if $L \succ L'$ 
then $L' \not\succ L$), and transitivity (i.e., if $L \succ L'$ and 
$L' \succ L''$ then $L \succ L''$) properties.
\end{assumption}
Since both feedback and preferences are generated from the same utility $u_x$, it is natural to assume that they are consistent:
\begin{assumption}[Feedback-preference consistency]
\label{ass:rating_consistency}
    For all slates $L, L' \in \mathcal{L}$ and any user $x \in \mathcal X$, ratings always align with user preferences. That is, if $L \succ_x L'$, then $\mathbb E[R | L] > \mathbb E[R | L']$.
\end{assumption}
It is common to interpret the expected feedback on a slate $L$ by a user $x$ as its utility score $u_x(L) = \mathbb E[R | L]$, which by definition is consistent.

\subsection{Ranking systems from preferences}
\label{sec:ranking}

It is common to compare systems using crowd-sourced user feedback (A/B testing) on their respective outputs~\cite{chiangChatbotArenaOpen2024b,ameli2025a}. When ranking from pairwise preferences, users are sequentially presented with pairs of outputs (one output from each system) and indicate their preference on each pair. The specific problem of ranking $C \geq 2$ RecSys candidates according to preferences can be formulated as follows.

Let $\mathcal A := \{ (a, a') : a, a' \in [C]\}$ denote the set of all combinations of RecSys pairs to rank. At each time $t$, a pair $A_t := (A_{t,1}, A_{t,2}) \in \mathcal A$ is selected, both RecSys produce slates for user $x_t$, and we observe preference $H_t \in \{0, 0.5, 1\}$, i.e., the slate $A_{t,1}$ was preferred; there was a tie; or the slate $A_{t,2}$ was preferred. Let $\boldsymbol \xi := (\xi_1, \ldots, \xi_C) \in \mathbb{R}^C$ denote the vector of latent strengths, where $\xi_a$ represents the expected preference for the slates generated by RecSys $a$: higher values mean that the recommendations from RecSys $a$ are more likely to be preferred in pairwise comparisons. The objective is to recover the true ranking as defined by the strengths, $\operatorname{rank}: \mathbb R^C \mapsto \operatorname{Sym}(C)$, where $\operatorname{Sym}(c)$ is the set of all permutations of $\{1, 2, \dots, c\}$.

Strengths can be estimated from the collected preferences $\{H_t\}_{t=1}^T$ using predefined statistical models. The common approach is to impose a statistical model on $P_{\boldsymbol{\theta}}(H_t | A_t)$ and consider the likelihood of its parameters $\boldsymbol \theta$. 

Let $\mathbf W=[w_{aa'}]$ and $\mathbf T=[t_{aa'}]$ count, respectively, the wins of $a$ over $a'$ and the ties, with $n_{aa'}:=w_{aa'}+w_{a'a}+t_{aa'}$ total comparisons (App.~\ref{app:rao-kupper}).
The likelihood of parameters $\boldsymbol \theta$ corresponds to the product of multinomial distributions for each pair $(a,a')\in\mathcal A$:
\begin{align}
\label{eq:likelihood}
    \mathcal L(\boldsymbol \theta| \mathbf W, \mathbf T) := \prod_{(a, a')} \kappa_{aa'}\prod_{h} P_{\boldsymbol{\theta}}(h | (a, a')),
\end{align}
where $\kappa_{aa'} := \frac{n_{aa'}!}{w_{aa'}! w_{a'a}! t_{aa'}!}$. More specifically, the estimated parameters are selected to maximize the logarithmic likelihood given the observed preferences: $\hat{\boldsymbol \theta} = \operatornamewithlimits{argmax}_{\boldsymbol \theta} \log \mathcal L(\boldsymbol \theta | \mathcal A, \mathbf W, \mathbf T)$. The estimated parameters $\hat{\boldsymbol \theta}$ yield the estimated strengths $\hat{\boldsymbol{\xi}} \in \mathbb R^C$ required to rank the RecSys agents, along with other parameters that govern the probability of outcomes. We present an experimental use case (Section~\ref{sec:use_case}) in which we instantiate the recent generalized Rao-Kupper model~\citep{ameli2025a} to rank several slate recommender systems.

\subsection{Offline evaluation}
\label{sec:offline-eval}

Collecting preferences from real users imposes substantial costs: experiments require millions of users to detect small effects, risk degrading user experience during exploration, and operate on timescales of weeks~\citep{gomez-uribeNetflixRecommenderSystem2015}.  In practice, evaluation is therefore a funnel: inexpensive offline estimates prune the candidate pool, and only the surviving few graduate to a live experiment. Our aim is to strengthen this offline stage and not to substitute for the final online one. We consider offline evaluation through a \emph{synthetic evaluator} $f : \mathcal L \times \mathcal L \times \mathcal X \mapsto \{ 0, 0.5, 1 \}$ that, given a user $x$ and two slates $L, L'$, returns $0$ if $L$ is preferred, $1$ if $L'$ is preferred, or $0.5$ for a tie (a stand-in for the human preference outcome $H_t$). The evaluator conditions on the user through their past items \emph{and} the feedback received on them: it rests on the collaborative-filtering assumption transposed to item semantics, i.e. a user who favoured semantically similar items will prefer slates consistent with that pattern.

Consider a dataset $\mathcal D = \{(x_n, \mathcal I_n, \mathcal R_n)\}_{n=1}^N$ of pre-collected feedback from $N$ users, where $\mathcal I_n = (i_{n,1}, i_{n,2}, \dots)  \subseteq \mathcal I$ is the candidate item pool of the $n$-th user and $\mathcal R_n = (r_{n,1}, r_{n,2}, \dots)$ contains the associated per-item interactions, i.e., the standard format for RecSys benchmarks \cite{Swaminathan2016OffpolicyEF}. Algorithm~\ref{alg:build_matrices} describes the offline pairwise preference collection procedure using a synthetic evaluator. 
For each user in the dataset, for each pair of RecSys, both candidates are queried to recommend a slate  
and the preference articulated by the synthetic evaluator feeds the count matrices required by statistical models (Section~\ref{sec:ranking}). 

\begin{algorithm}
\caption{Synthetic A/B testing}
\label{alg:build_matrices}
\begin{algorithmic}[1]
\Require Dataset $\mathcal D = \{(x_n, \mathcal I_n, \mathcal R_n)\}_{n=1}^N$, possible slates $\mathcal L$, and synthetic evaluator $f$
\State $\mathbf{W}, \mathbf{T} \gets \mathbf{0}_{C \times C}$
\For{$n \in [N]$, $(a, a') \in \binom{[C]}{2}$}
    \State $(L, L') \gets a(\mathcal L), a'(\mathcal L)$
    \State $H \gets f(L, L' \mid x_n)$
    \State $w_{aa'} \mathrel{+}= \mathbf{1}\{H{=}0\}$
    \State $w_{a'a} \mathrel{+}= \mathbf{1}\{H{=}1\}$
    \State $t_{aa'}, t_{a'a} \mathrel{+}= \mathbf{1}\{H{=}0.5\}$
\EndFor
\State \Return $\mathbf{W}, \mathbf{T}$
\end{algorithmic}
\end{algorithm}


\section{Validating synthetic preferences}
\label{sec:validation_protocol}

The proposed offline evaluation (Algorithm~\ref{alg:build_matrices}) is only useful if the preferences articulated by the synthetic evaluator are reliable. We therefore introduce a validation protocol to assess synthetic preferences using a dataset $\mathcal D = \{(x_n, \mathcal I_n, \mathcal R_n)\}_{n=1}^N$, assuming a pool of at least $K$ candidate items for each of the $N$ users, i.e., $|\mathcal I_n| \geq K~\forall n\in [N]$. The set of slate candidates for the $n$-th user, $\mathcal L_n \subseteq \mathcal L$, denotes all possible ordered sequences of $K$ distinct items from $\mathcal I_n$. We evaluate external alignment with empirical user utilities and internal logical coherence.


\subsection{\textbf{External coherence}}
\label{sec:external_coherence}

External coherence measures the alignment between a synthetic evaluator and empirical preferences estimated from the dataset. This first requires defining \emph{slate-level} metrics using the available \emph{item-level} information ($\mathcal I_n$, $\mathcal R_n$). This is not a circular check: the evaluator sees only rendered text, never the empirical utility nor the held-out interactions and popularity statistics used to compute it, so external coherence compares two signals with disjoint inputs. Moreover, RecSys traditionally use such indicators \emph{for learning}, where the bias enters the model \citep{chen2023bias}; we use them for evaluation only.


\paragraph{From per-item interactions to slate-level metrics}

Let $\omega_{\mathcal D, n} : \mathcal L \to \mathbb R$ denote a scoring  metric that aggregates the per-item interactions of the $n$-th user in dataset~$\mathcal D$ into a score for any slate $L\in\mathcal L$. 
Several metrics have been proposed to capture various aspects of user utility \cite{zhao2025Fairness}. \textit{Normalized discounted cumulative gain} (nDCG) scores a slate $L$ based on logged user-interactions ($\mathcal R_n$) with items $i\in L$, weighted by their position in the slate~\citep{jarvelin2002cumulated}. The ideal slate w.r.t. nDCG for a given user contains the $K$ items with the most interactions for that user, in decreasing order of interactions. 
Let the set of \textit{expected} items $\mathcal E_n \subset \mathcal I_n$ denote the top-$K$ items (according to $\mathcal R_n$) for the $n$-th user. \textit{Serendipity} scores a slate $L$ based on user-interactions ($\mathcal R_n$) with unexpected items $i\in L, i\in \mathcal I_n, i\not\in \mathcal E_n$~\citep{kaminskas2017diversity}. Let $p(i) := |\sum_{n=1}^N \mathbf 1\{ i\in\mathcal I_n \}|/N$ denote the fraction of users who have been exposed to item $i$. \textit{Novelty} scores a slate $L$ based on the rarity $-\log_2 p(i)$ of all items $i\in L$~\citep{vargas2011rank}. Finally, \textit{diversity} scores a slate $L$ based on the Jaccard dissimilarity between all pairs of items $i,j\in L$ ~\citep{kaminskas2017diversity}. 
See Appendix~\ref{app:external-metrics} for complete formal definitions.

\paragraph{From metrics to preferences}
Following prior work, we aggregate several metrics using linear scalarization~\citep{lin2019pareto} into a single scalar value that approximates user utility.
\begin{definition}[Empirical utility]
\label{def:empirical_utility}
Let $\Omega$ denote a set of metrics. For any metric $\omega\in\Omega$, let $\omega_{\mathcal D, n}(L)$ denote the metric $\omega$ applied to slate $L\in\mathcal L$ for the $n$-th user in dataset $\mathcal D$ and let $\bar{\omega}_{\mathcal D, n}(L)$ denote its $[0,1]$-rescaling (on $\mathcal D$). The utility of slate $L\in\mathcal L$ for the $n$-th user in dataset $\mathcal D$ is defined as
\begin{align*}
    \tilde u_{\mathcal D, n}(L) := \sum_{\omega \in \Omega} \alpha_\omega \cdot \bar{\omega}_{\mathcal D, n}(L),
\end{align*}
where the scalarization weights $\boldsymbol \alpha = (\alpha_\omega)_{\omega \in \Omega}$ satisfy $\alpha_\omega \geq 0$ and $\sum_{\omega \in \Omega} \alpha_\omega = 1$.
\end{definition}
The weights $\boldsymbol \alpha$ can be set to reflect domain-specific priorities (e.g., emphasizing diversity for music playlists), learned from A/B testing, or treated as hyperparameters subject to sensitivity analysis.
By the definition of preferences (Def.~\ref{def:pref}), the slate with the highest utility should always be preferred in a pairwise comparison. The \textit{utility gap} captures the direction that the preference should follow according to the utility scores:
\begin{align}
\label{eq:utility_gap}
\Delta_{\mathcal D, n}(L, L') := \tilde u_{\mathcal D, n}(L) - \tilde u_{\mathcal D, n}(L').
\end{align}
Given two slates $L, L'\in\mathcal L$, we estimate the preference of the $n$-th user in dataset $\mathcal D$ based on the empirical utility gap (Eq.~\ref{eq:utility_gap}):
\begin{align}
\label{eq:empirical_pref}
y_{\mathcal D, n}(L, L') := 
\begin{cases}
    0 & \text{if } \Delta_{\mathcal D, n} > 0\\ 
    0.5 & \text{if }  \Delta_{\mathcal D, n} = 0\\ 
    1 & \text{if }  \Delta_{\mathcal D, n} < 0. 
\end{cases}
\end{align}
%
We compare the preferences of a synthetic evaluator with the empirical preferences (Eq.~\ref{eq:empirical_pref}).
\begin{definition}[External accuracy]
\label{def:ext_accuracy}
The external accuracy of a synthetic evaluator $f$ w.r.t. the $n$-th user in dataset $\mathcal D$ is given by
\begin{align*}
    \operatorname{Acc}_{\mathcal{D}, n}(f) := \hspace{-15pt}
    \sum_{(L,L') \in \mathcal{L}_n^2} \hspace{-10pt}
    \frac{\mathbf{1}\{f(L,L'|x_n) = y_{\mathcal D, n}(L,L')\}}{|\mathcal{L}_n^2|}.
\end{align*}
The global accuracy averages over all users:
\begin{align*}
    \overline{\operatorname{Acc}}_{\mathcal{D}}(f) := 
    \frac{1}{N} \sum_{n=1}^{N} 
    \operatorname{Acc}_{\mathcal{D}, n}(f).
\end{align*}
\end{definition}

\subsection{\textbf{Internal coherence}}
 
Internal coherence measures the logical consistency of synthetic preferences by testing satisfaction of the preference axioms (Assumption~\ref{ass:pref_props}). 
Prior work \citep{zhaoMeasuringInconsistencyLarge2024a,liu2025aligning, wang2026trustjudge} applied these axioms to expose inconsistencies in LLM preferences over individual items and text responses. We apply them to slates, where the inter-item interactions (in a given slate) introduce challenges not captured by previous benchmarks.
We estimate one score per axiom over the set of users and their slate candidates in the dataset $\mathcal D$. 
 
\begin{definition}[Irreflexivity score]
\label{def:irreflexivity_score}
Self-comparisons should yield indifference:
\begin{align*}
    \operatorname{Irr}_{\mathcal D}(f) := \sum_{n=1}^N \sum_{L \in \mathcal L_n} \frac{\mathbf{1}\{f(L, L|x_n) = 0.5\}}{N |\mathcal L_n|} .
\end{align*}
\end{definition}
 
\begin{definition}[Asymmetry score]
\label{def:asymmetry_score}
For $\mathcal{P}_n = \{(L, L') \in \mathcal L_n^2 : f(L, L'|x_n) = 0\}$, reversing the comparison order should flip the preference:
\begin{align*}
    \operatorname{Asy}_{\mathcal D}(f) := \sum_{n=1}^N \sum_{(L, L') \in \mathcal{P}_n} \hspace{-10pt} \frac{\mathbf{1}\{f(L', L|x_n) = 1\}}{N |\mathcal{P}_n|} .
\end{align*}
\end{definition}
 
\begin{definition}[Transitivity score]
\label{def:transitivity_score}
Let $\mathcal{T}_n = \{(L, L', L'') \in \mathcal L_n^3 : f(L, L'|x_n) = 0 \wedge f(L', L''|x_n) = 0\}$ be the set of all triples on which $f$ asserts the preference chain $L \succ_{x_n} L' \succ_{x_n} L''$. The transitively implied preference on the endpoints, $L \succ_{x_n} L''$, should hold:
\begin{align*}
    \operatorname{Tra}_{\mathcal D}(f) := \sum_{n=1}^N  \sum_{(L, L', L'') \in \mathcal{T}_n} \hspace{-15pt} \frac{\mathbf{1}\{f(L, L''|x_n) = 0\}}{N |\mathcal{T}_n|}.
\end{align*}
\end{definition}
 
All metrics operate purely via inference-time queries and require no access to model internals.

\section{Empirical Study}
\label{sec:empirical}


This section establishes LLMs as practical evaluators of slate recommendations through a sequence of controlled experiments using the proposed validation protocol (Section~\ref{sec:validation_protocol}). We leverage the in-context learning capabilities of LLMs to generate synthetic preferences over slates~\citep{dong-etal-2024-survey}: given a user $x \in \mathcal{X}$ and two slates $L,L'\in\mathcal L$, the LLM is prompted to determine which slate is better aligned with the preference of user $x$. 
Figure~\ref{fig:eval_setting} illustrates the proposed setting. See Appendix~\ref{app:prompts} for prompt details. 

\begin{figure}[t]
  \centering
  \includegraphics[
    width=\columnwidth  
  ]{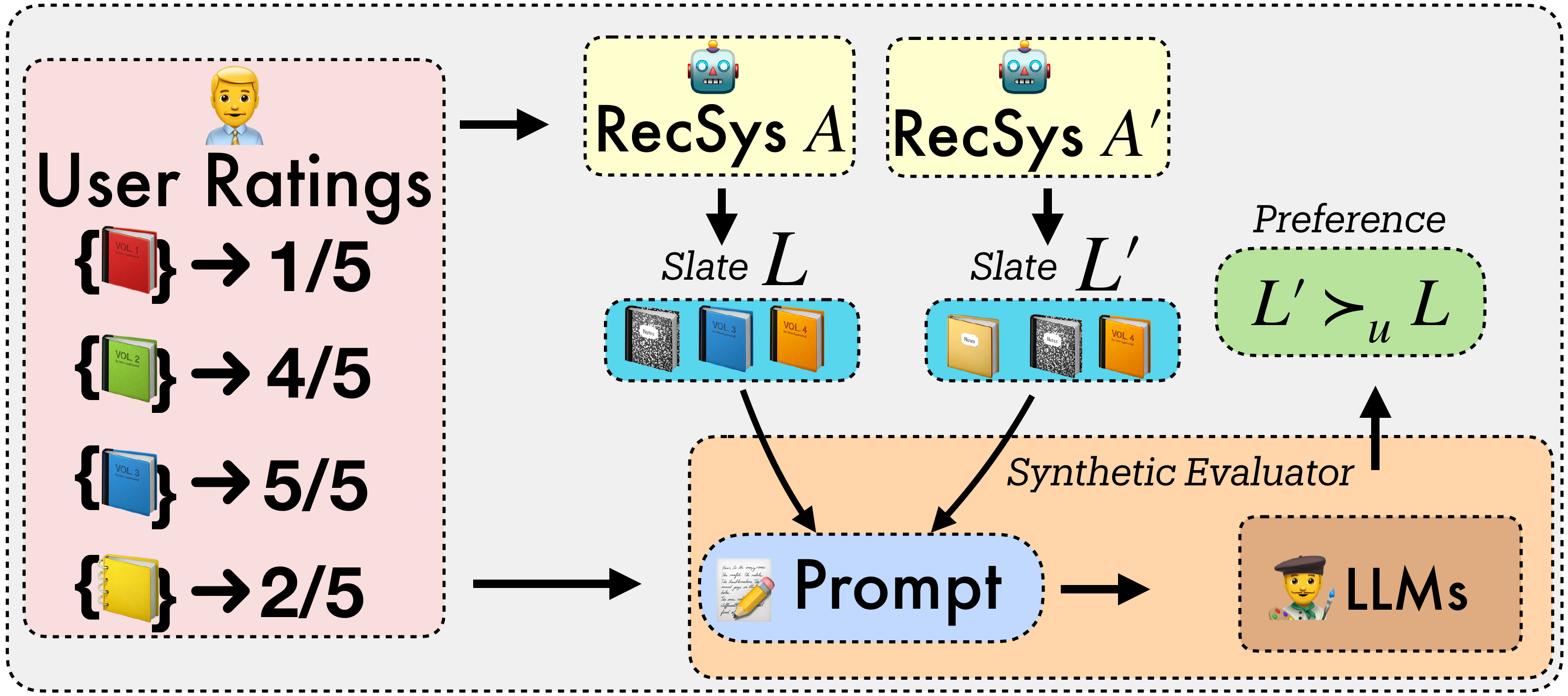}
  \caption{Overview of LLM-based A/B testing.}
  \label{fig:eval_setting}
\end{figure}


\paragraph{Tasks and datasets}

We validate on the most realistic slate problem of \emph{joint selection and ordering}, in which a RecSys must decide \emph{which items} to recommend \emph{and} \emph{how} to order them. We conduct experiments on two datasets:
\textsc{Spotify} \cite{chenRecsysChallenge20182018} (user-defined playlist orders) and \textsc{MIND} \citep{wuMINDLargescaleDataset2020} (click chronologies). All main results are reported at slate size $K = 10$ sensitivity to the operating point ($K \in \{2, 5\}$)  is analysed in Appendix~\ref{app:auxiliary-tasks}. 
The composite empirical utility (Def.~\ref{def:empirical_utility}) uses an \emph{nDCG-centered} scalarization $\boldsymbol \alpha$, the dominant objective in deployed pipelines; Appendix~\ref{app:weight_regimes} re-runs the study under two contrasting weight regimes and shows the LLM ordering is preserved.
Complete dataset statistics, along with a random baseline and results on ablation tasks (\emph{item selection only} and \emph{ordering-only}) are deferred to App.~\ref{app:datasets} and~\ref{app:auxiliary-tasks}.
 

\paragraph{Benchmarked configurations}

We consider three LLM categories: Qwen2.5~\cite{qwen2025technicalreport}, DeepSeek~\cite{deepseekai2024llms}, and Mistral~\cite{liu2026ministral3}. For Qwen2.5 and DeepSeek, we consider three specializations (base/general, coder, and math) which share the same base architecture but differ in pre-training, enabling a controlled study of specialization effects within a single model family. We evaluate a basic single-query approach and explore the potential of ensembling designs with two diversity-enhancing mechanisms, resulting in 35 configurations (see details in the following subsections).

\paragraph{Overview of results}

\begin{figure*}
  \centering
  \includegraphics[width=\textwidth]{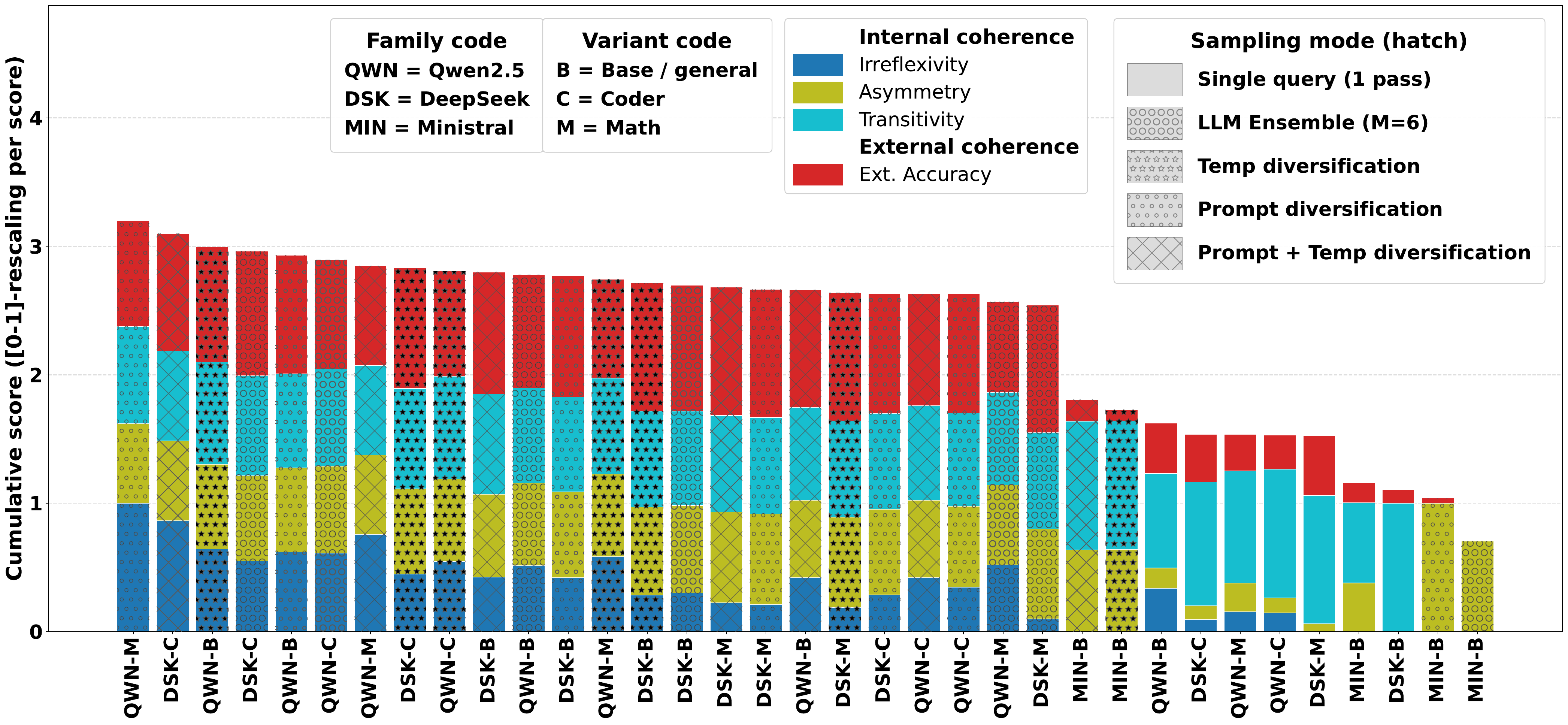}
  \caption{External and internal coherence averaged over both datasets for all configurations. }
  \label{fig:axiom_stacked_bars_waccuracy}
\end{figure*}

Figure~\ref{fig:axiom_stacked_bars_waccuracy} shows the external and internal coherence for each configuration, averaged over the two datasets, 
sorted in decreasing order of cumulative score. 
\textbf{Key obs. 1)} Ensembling ($M=6$, hatches \texttt{o} and \texttt{x}) systematically beats single-query (no hatch), which collapses at the bottom: every model gains $\ge 1.0$ cumulative point. 
\textbf{Key obs. 2)} Among ensembling modes, temperature diversification (\texttt{*}) provides the largest gain, above prompt diversification and both-sampling while they remain competitive. \textbf{Key obs. 3)} Qwen2.5-Math dominates the top cumulative rank, supporting the view that math training improves logical consistency ~\citep{guo2025deepseek}. However, Qwen2.5-Base is the second-best configuration overall and outperforms its coder variant, meaning the math/code advantage is family-specific (clear on Qwen2.5-Math, marginal on Qwen2.5-Coder) and modulated by the diversification strategy. \textbf{Key obs. 4)} DeepSeek-Math attains the highest raw external accuracy (0.66) but drops to mid-rank in cumulative score because its irreflexivity is near zero (it readily declares a slate worse than itself), which makes external alignment alone an incomplete proxy for preference articulation and motivates the joint reporting of both axes.


\subsection{Single Query}
\label{sec:single-query}

The simplest off-the-shelf protocol is to implement the synthetic preference ($f$) using a single LLM query where, given user $x$ and two slates $L, L'$, the LLM outputs $L$ (0), $L'$ (1), or tie (0.5). LLMs are known to suffer from \emph{positional bias}, flipping the verdict on input-order reversal, and \emph{neutral-label bias}, over-using ``tie'' as an escape hatch in borderline cases~\citep{zhengJudgingLLMasaJudgeMTBench2023}. Ensembling designs are investigated to avoid these problems. 

\subsection{Ensembling}
\label{sec:llm-ensemble}
 
Allocating additional inference-time compute to a fixed model is a well-established strategy to improve reliability~\citep{snell2025scaling}. We instantiate this principle for preference elicitation: an LLM ensemble aggregates the responses of $M$ LLM instances ($M$ even) using a voting mechanism, trading a higher query budget for more robust synthetic preferences. To prevent neutral-label bias, we consider a design in which each instance $m\in[M]$ is queried with a binary choice (no tie option):

\begin{align*}
\operatorname{LLM}^{(m)}(L, L' | x) := \begin{cases}0 & \text{if } m \text{ prefers } L\\ 1 & \text{if } m \text{ prefers } L'.\end{cases}
\end{align*}
The queries are also designed to control for position bias: when comparing slates $L$ against $L'$, half of the instances are queried with $(L, L')$, and the other half are queried with $(L', L)$. We count the votes in favor of slate $L$ in both orders:
\begin{align}
\label{eq:vote_count}
v_{L,L'}(x) &:= \sum_{m=1}^{M/2} \mathbf{1}[\operatorname{LLM}^{(m)}(L,L'|x) = 0] \\
&\quad + \sum_{m=M/2+1}^{M} \mathbf{1}[\operatorname{LLM}^{(m)}(L',L|x) = 1]\nonumber
\end{align}
and let the synthetic preference ($f$) be the majority:
\begin{align*}
\widehat{\operatorname{LLM}}_M(L, L'|x) := \begin{cases}0 & v_{L,L'}(x) > M/2\\ 1 & v_{L,L'}(x) < M/2\\ 0.5 & v_{L,L'}(x) = M/2.\end{cases}
\end{align*}


\paragraph{Ensemble size}
A detailed analysis of ensemble-size effects is provided in Appendix~\ref{app:votes} (Figs.~\ref{fig:vote_accuracy_by_task}). Accuracy improves non-monotonically with $M$ but plateaus near $M{=}6$ (Fig.~\ref{fig:vote_accuracy_by_task}); we use $M{=}6$ throughout, balancing resolution and cost.

\begin{figure}[t]
  \centering
  \includegraphics[width=0.8\columnwidth]{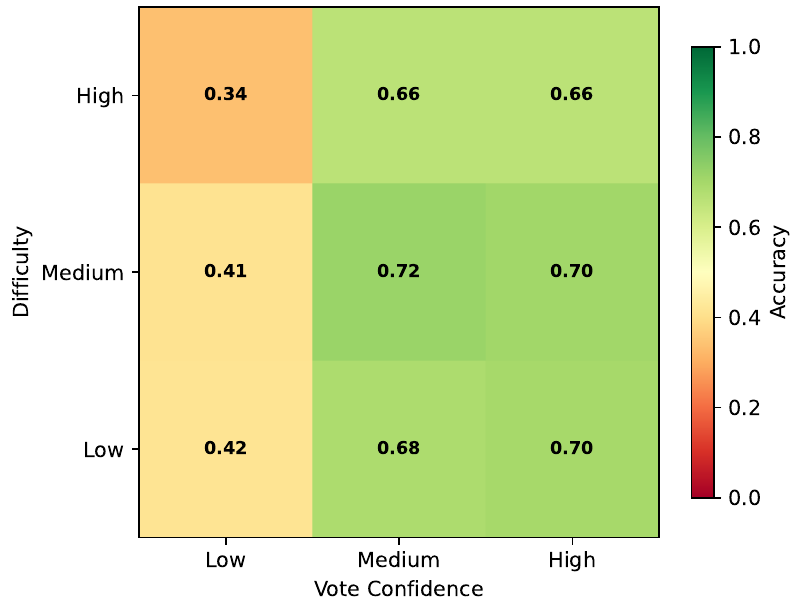}
  \caption{Pairwise accuracy stratified by difficulty level and vote confidence $\delta_M$ (Eq.~\ref{eq:vote_confidence}). High-confidence verdicts are more accurate at every difficulty level.}
  \label{fig:confidence_difficulty_heatmap}
\end{figure}

\paragraph{Confidence margin}
The vote count (Eq.~\ref{eq:vote_count}) induces a per-comparison reliability score:
\begin{align}\label{eq:vote_confidence}
\delta_M(L, L' | x) := \frac{| 2\, v_{L,L'}(x) - M |}{M} \in [0, 1],
\end{align}
with $\delta_M{\to}1$ for unanimous votes and $\delta_M{\to}0$ for near-ties. A perfect $M/2$ split ($\delta_M{\to}0$) signals a genuine semantic ambiguity.
Figure~\ref{fig:confidence_difficulty_heatmap} reports pairwise accuracy jointly stratified by vote confidence $\delta_M$ and comparison difficulty (based on slate similarities, see App.~\ref{app:datasets}). Empirically, $\delta_M$ tracks the ensemble's correctness rate: high-confidence verdicts are more accurate at every difficulty level. 


Table~\ref{tab:spearman_external}  (App.~\ref{app:auxiliary-tasks}) reports the spearman correlations, pooled over every (user, slate-pair), 
between the per-pair absolute utility gap $|\Delta_{\mathcal D, n}(L, L')|$ (Eq.~\ref{eq:utility_gap}) and the ensemble ($M=6$) correctness indicator $\mathbf{1}\{f(L,L'|x_n) = y_{\mathcal D, n}(L,L')\}$ (Def.~\ref{def:ext_accuracy}), denoted $\rho_{\operatorname{Acc}}$.
Correctness is consistently and strongly anti-correlated with the gap, $\rho_{\operatorname{Acc}} \in [-0.69, -0.50]$, $p < 0.001$ for all family-variants on both datasets: judgment accuracy varies in a gap-sensitive manner, the calibration signature of a reliable evaluator. 
Similar results are observed on auxiliary tasks (Table~\ref{tab:spearman_external_full}, App.~\ref{app:auxiliary-tasks}).
 

\subsection{Diversity enhancement}
\label{sec:diversity}


So far, the $M$ instances differ only by input order. We investigate two complementary mechanisms to increase the ensemble diversity. With \emph{prompt diversification}~\citep{zhou2023large}, we consider a set of prompt variants $\mathcal{V}$ and assign the variant at index $(m{-}1) \bmod |\mathcal{V}|$ to the $m$-th instance (Appendix~\ref{app:prompts}). We also apply \emph{decoding diversification} \citep{snell2025scaling} by assigning each instance $m\in [M]$ a symmetric temperature schedule around $T^{\text{base}} = 0.7$:
\begin{align*}
T^{(m)} := \max\!\left(0.05,\; T^{\text{base}} + 0.2\!\left(m - \tfrac{M}{2} - \tfrac{1}{2}\right)\right).
\end{align*}

\section{Use case: Ranking recommender systems from synthetic preferences}
\label{sec:use_case}


We evaluate whether the LLM-based ranking recovers a meaningful ordering over different RecSys. We consider six architecturally diverse strategies: LightGCN~\citep{heLight2020},  SASRec~\citep{WangSASrec2018}, BPR~\citep{rendleBPRBayesianPersonalized2009}, GRU4Rec~\citep{HidasiGRU4Rec2018}, MultiVAE~\citep{LiangMultiVAE2018} and 
DeepFM~\citep{guoDeepFM2017}. These span matrix factorization, graph convolution, sequence modelling, and feature interaction, providing a broad and theoretically well-separated testbed.
We use the same tasks and datasets as those introduced in Section \ref{sec:empirical}.



\paragraph{Ranking evaluation}


For each RecSys $a \in [C]$, let $L^{(n)}_a$ denote the slate recommended to user $n$. We compute the \textit{empirical strength} $\tilde \xi(a)$ of RecSys $a$ on dataset $\mathcal D$ as the average empirical utility (Def.~\ref{def:empirical_utility}) of the slates recommended by $a$ for each of the $N$ users contained in the dataset:
\begin{equation*}
    \tilde \xi(a) := \frac{1}{N} \sum_{n=1}^N \tilde u_{\mathcal D, n}\big(L^{(n)}_a\big).
\end{equation*}
We denote the resulting \textit{empirical strength vector} as $\tilde{\boldsymbol \xi} \in \mathbb R^C$. 
We report the rank-correlation Kendall~$\tau$ between $\operatorname{rank}(\tilde{\boldsymbol \xi})$ and the rankings retrieved by the benchmarked methods (Section~\ref{sec:ranking}). 

\paragraph{Ranking methods} We compare the rankings obtained using the LLM-based synthetic A/B testing against two off-policy estimators (OPE). The first is inverse propensity scoring (IPS)~\cite{liUnbiasedOfflineEvaluation2011}, the standard OPE for RecSys~\citep{GilotteOffline2018}: given a dataset $\mathcal D$, IPS estimates item propensities and scores each slate via a propensity-weighted nDCG. The second is the pseudoinverse estimator (PI)~\citep{Swaminathan2016OffpolicyEF}, a slate-specific OPE that reduces variance in the combinatorial slate space and can score slates absent from the dataset. For both estimators, the resulting scores are converted into pairwise win/tie matrices.
%
The LLM-based approach uses ensembling ($M = 6$) with enhanced diversity via prompt diversification as a synthetic evaluator (Alg.~\ref{alg:build_matrices}). Win/tie matrices are fed into the same statistical model to generate the estimated strengths vectors $\hat{\boldsymbol\xi}^{\text{IPS}}$, $\hat{\boldsymbol\xi}^{\text{PI}}$, and $\hat{\boldsymbol\xi}^{\text{LLM}}$, respectively (Section~\ref{sec:ranking}).

\paragraph{Statistical model}
 
We rank the RecSys candidates using the generalized Rao-Kupper model~\cite{ameli2025a}, described in App.~\ref{app:rao-kupper}. 
%
Maximum-likelihood fitting (Eq.~\ref{eq:likelihood}) yields the strengths $\hat{\boldsymbol \xi} \in \mathbb R^C$, a normalized score-difference matrix $Z = [z_{aa'}]$ encoding score gaps and uncertainty, and marginal confidence intervals $\mathcal C_a$ from a sandwich variance estimator~\citep{huber1967behavior}.
The approximate rank of RecSys~$a$, $\operatorname{rank}(\hat{\boldsymbol \xi})_a$, is obtained by comparing its marginal confidence interval with those of other candidates (Eq.~\ref{eq:approx_rank}, App.~\ref{app:rao-kupper}). 

\subsection{Results}

\begin{figure}[h]
  \centering
  \includegraphics[width=0.9\columnwidth]{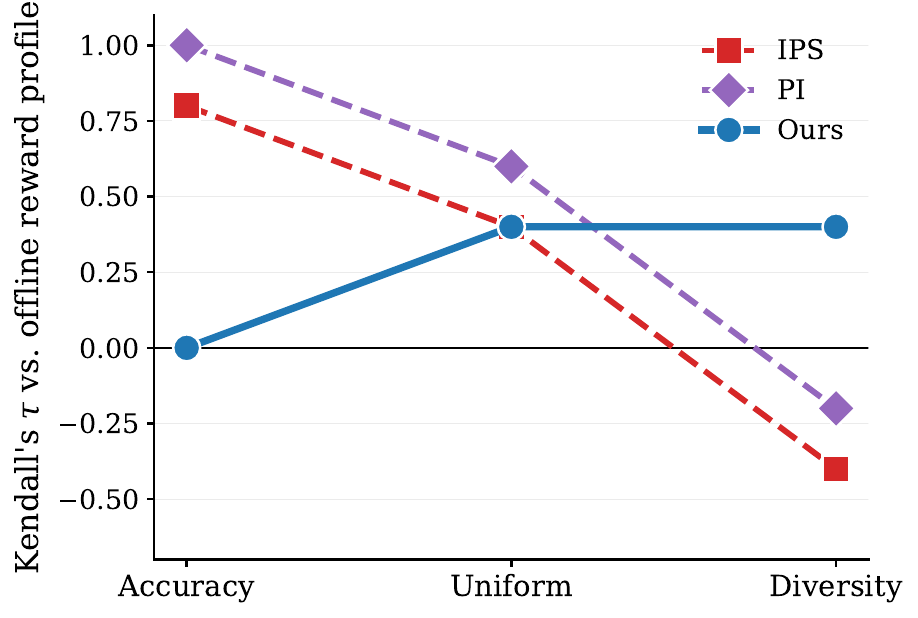}
  \caption{RecSys Eval. Kendall's $\tau$ vs offline reward}
  \label{fig:kendall_tau_gt_task3}
\end{figure}

Figure~\ref{fig:kendall_tau_gt_task3} reports, for each judge, the Kendall $\tau$ between its recovered ranking of the $C=6$ base RecSys candidates and the rank induced by a scalarized ground-truth utility $\tilde{\boldsymbol{\xi}}$, swept over three representative reward profiles (\emph{Accuracy}, \emph{Uniform}, \emph{Diversity}; details in App.~\ref{app:weight_regimes}). IPS and PI agree closely on \emph{Accuracy} ($\tau = +0.80$ and $+1.00$) but invert on \emph{Diversity} ($\tau = -0.40$ and $-0.20$), a swing of $1.20$ — a direct consequence of reusing historical exposure logs dominated by popularity-driven slates. The LLM-based judge is the only one that remains stable across profiles ($\tau \in \{0.00, +0.40, +0.40\}$, swing $0.40$). Absolute $\tau$ values do not reach significance ($p \in [0.08, 1.00]$), as expected with only $\binom{6}{2}=15$ pairs; the takeaway is that logged-data baselines are reliable only when the target reward is essentially accuracy (itself a known-incomplete proxy for recommendation quality \citep{mcnee2006accurate, ge2010beyond, zhou2010solving}), while the LLM-based judge delivers a stable ranking without requiring any logged propensity data. Extending the arena to the five training configurations of each base recommender ($C=30$), where rank correlations become statistically testable, confirms this pattern: the off-policy estimators' sign inversion under \emph{Diversity} reaches significance ($p=0.010$ on \textsc{MIND}, $p=0.020$ on \textsc{Spotify}), while a paired user-level bootstrap ($1{,}000$ resamples) shows the LLM judge's cross-profile swing is significantly smaller, its $\tau$ values staying of the same order across regimes.

\begin{figure}[h]
  \centering
  \includegraphics[width=0.9\columnwidth]{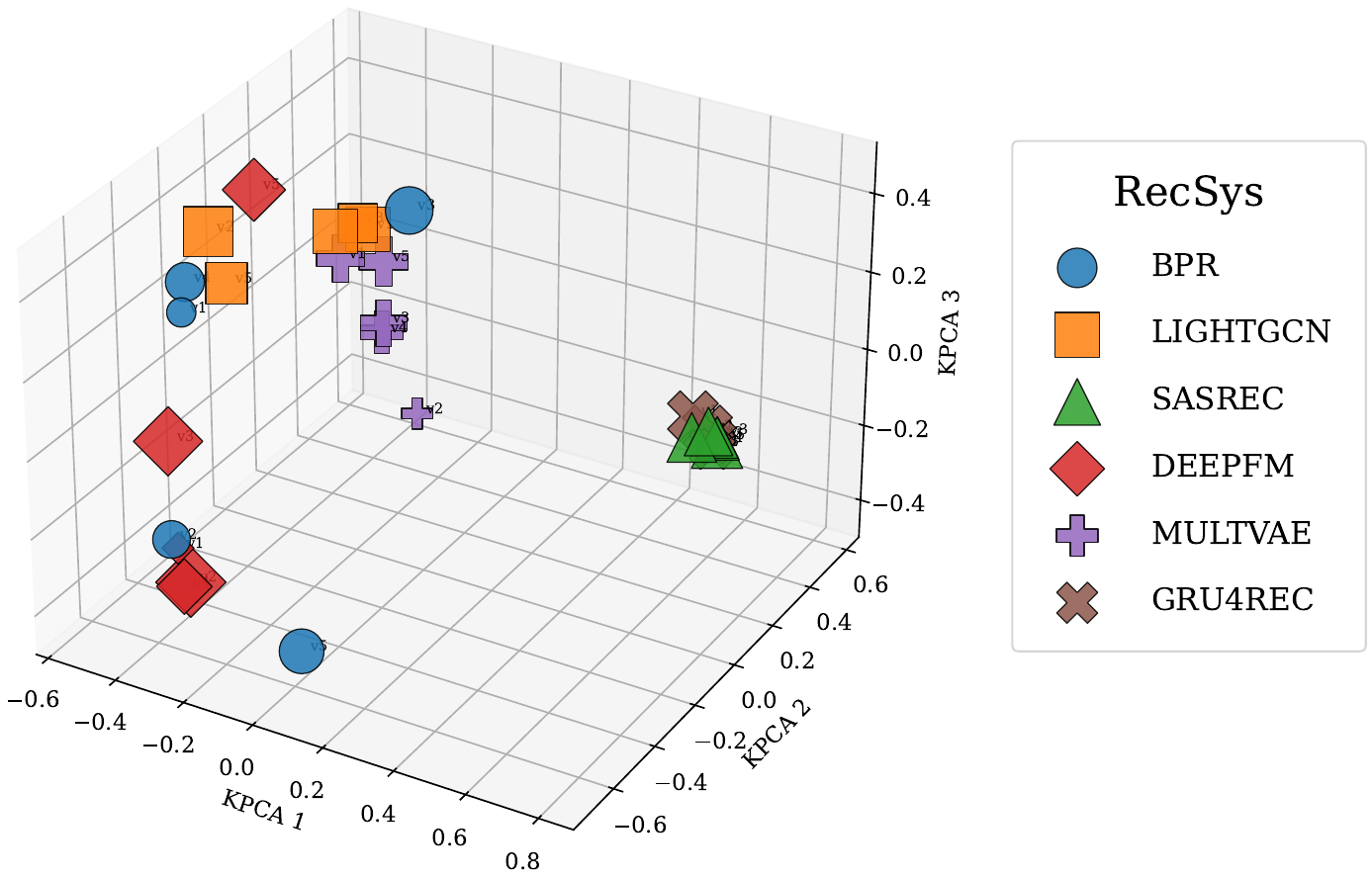}
  \caption{Offline-metric KPCA of RecSys candidates}
  \label{fig:ranking_recsys}
\end{figure}

\paragraph{Geometric structure of RecSys similarities}
Beyond ordinal ranking, our framework produces a full pairwise dissimilarity matrix $Z = [z_{aa'}]$ (Eq.~\ref{eq:normalized_z}), where each entry encodes both the score difference and the estimation  uncertainty between a pair of candidates. This matrix can be projected 
onto a low-dimensional space via Kernel PCA, yielding a geometric representation of the RecSys landscape (Figure~\ref{fig:ranking_recsys}). 

Distances reflect how confidently two candidates can be distinguished: distant pairs produce systematically different slates, while nearby ones are harder to discriminate and may be interchangeable in practice. This output (which is unavailable from IPS and PI) lets practitioners drop redundant candidates and prioritize informative comparisons.

\section{Conclusion}
We set out to test whether LLMs are promising to serve as offline evaluators of slate recommender systems 
and our experiments support an affirmative answer. We devised a validation protocol to evaluate external coherence of LLM-articulated preferences using empirical (pre-collected) data and internal coherence with preference axioms. We benchmarked LLM-based variants from different families and found that LLMs provide stable ranking regardless of which weighting of offline metrics is given, whereas IPS/PI do not.


\section{Limitations} 

Our external coherence protocol validates LLM preferences against a \emph{pre-collected} empirical utility derived from logged interaction data; it does not constitute a substitute for a live A/B test. Before the community can adopt LLM-based offline evaluation as a drop-in replacement for human preference studies, several further validations are needed: running a true online A/B experiment alongside the LLM evaluation pipeline and measuring how faithfully synthetic preferences recover the human-observed ranking of RecSys candidates. Our work establishes that LLM preferences are \emph{internally consistent} and \emph{externally aligned with logged utility signals}, which is a necessary but not sufficient condition for this stronger claim. Beyond this, a large fraction of deployed RecSys pipelines represent items as opaque identifiers rather than as natural-language descriptions: the model receives item IDs, tabular feature vectors, and implicit feedback signals (click-through rates, dwell times, purchase events) rather than textual metadata. Our approach requires that slates be renderable as human-readable text so that an LLM can reason about their content; it therefore does not apply, without additional item-featurization work, to purely ID-based or embedding-based pipelines where no semantic description of the items is available. We searched for public data that could stand in for such a study (logged production A/B tests \citep{saito2020open} and randomized-exposure benchmarks \citep{gao2022kuairand}) but these are item or position-level with anonymised features, or lack slate-level pairwise outcomes


\bibliography{biblio_new}

\appendix

\section{External Coherence Metrics}
\label{app:external-metrics}
 
This appendix provides the formal definitions of the four metrics considered to evaluate external coherence (Section~\ref{sec:external_coherence}). Recall that a dataset $\mathcal D = \{ (x_n, \mathcal I_n, \mathcal R_n) \}_{n=1}^N$ contains pre-collected feedback from $N$ users, where $\mathcal I_n = (i_{n,1}, i_{n,2}, \dots) \subset \mathcal I$ is the candidate item pool for the $n$-th user and $\mathcal R_n = (r_{n,1}, r_{n, 2}, \dots)$ contains the associated per-item interactions. Let $k_n(i)$ denote the position (index) of item $i\in\mathcal I_n$, with $k_n(i) = 0$ if $i\not\in\mathcal I_n$. Then $r_{n,k_n(i)}$ denotes the per-interaction of item $i$ for the $n$-th user, with $r_{n,0} = 0$ if this user has never interacted with item, i.e., $i\not\in\mathcal I_n$. Assuming at least $K$ candidate items for each of the $N$ users, i.e., $|\mathcal I_n| \geq K~ \forall n\in [N]$, the set of slate candidates for the $n$-th user, $\mathcal L_n \subseteq \mathcal L$, denotes all possible ordered sequences of $K$ distinct items from $\mathcal I_n$. We use the notation $L = (i_1, i_2, \dots, i_K)$ to refer to the $K$ items of a slate.


\paragraph{nDCG} This metric aggregates per-item interactions based on their position in the slate, capturing both the orderings of recommended items and their relevance~\citep{jarvelin2002cumulated}:
\begin{align*}
\operatorname{nDCG}_n(L)
&:= \frac{\operatorname{DCG}_n(L)}
{\operatorname{DCG}_n(L^\star_n)}
\\
\text{with } \operatorname{DCG}_n(L)
&:= \sum_{j=1}^{K}
\frac{r_{n, k_n(i_j)}}{\log_2(j+1)},
\end{align*}
where $L^\star_n \in \mathcal L_n$ is the ideal slate for the $n$-th user, which contains the top-$K$ items $i\in\mathcal I_n$ sorted in decreasing order of interactions $r_{n,k_n(i)}$. 

\paragraph{Serendipity} Let the set of expected items $\mathcal{E}_n \subset \mathcal I_n$ for the $n$-th user denote the top-$K$ items with the highest number of user-interactions ($\mathcal R_n$). Serendipity aggregates per-item interactions of unexpected items in the slate~\citep{kaminskas2017diversity}:
\begin{align*}
    \operatorname{Ser}_n(L) := \frac{1}{K} \sum_{j=1}^{K} r_{n, k_n(i_j)} \cdot \mathbbm{1}\big[i_j \notin \mathcal{E}_n\big].
\end{align*}
An item contributes to the slate-level score only if it is has many interactions (high-relevance), but not too many (unexpected).

\paragraph{Novelty} Let $p(i) := \sum_{n=1}^N \mathbf 1 \{ i \in \mathcal I_n \} / N$ denote the fraction of users who have been exposed to item~$i$. Novelty aggregates the item-level rarity of the slate components into a slate-level score~\citep{vargas2011rank}: 
\begin{align*}
    \operatorname{Nov}(L) := \frac{1}{K} \sum_{j=1}^{K} -\log_2 p(i_j).
\end{align*}
Novelty is user-independent: it captures dataset-wide item rarity rather than user-specific unfamiliarity, serving as a proxy for long-tail exposure.

\paragraph{Diversity} Let $\mathcal{G}_i$ denote the set of attributes of item $i$ (e.g., genres, categories) available from the catalogue $\mathcal I$, and let $\operatorname{sim}(i, i') := |\mathcal{G}_i \cap \mathcal{G}_{i'}| / |\mathcal{G}_i \cup \mathcal{G}_{i'}|$ denote the Jaccard similarity. Diversity aggregates \emph{pairwise} item-level dissimilarities into a slate-level score~\citep{kaminskas2017diversity}:
\begin{align*}
    \operatorname{Div}(L) := \frac{1}{\binom{K}{2}} \sum_{1 \leq j < j' \leq K} \big(1 - \operatorname{sim}(i_j, i_{j'})\big).
\end{align*}
Diversity is computed directly from the slate $L$ and the catalogue metadata, independently of user expositions/interactions.

\section{Generalized Rao-Kupper Model}
\label{app:rao-kupper}
 
This appendix provides the complete specification of the generalized Rao-Kupper model~\citep{ameli2025a} used in Section~\ref{sec:use_case} to rank RecSys candidates from synthetic pairwise preferences (see Sections~\ref{sec:ranking} and~\ref{sec:offline-eval}). We reproduce the model here for completeness; the derivation and identifiability analysis are given in~\citet{ameli2025a}.
 
\paragraph{Thurstonian representation}
The generalized Rao-Kupper model adopts the Thurstonian representation~\citep{Thurstone1927}, in which unknown strengths $\boldsymbol \xi$ of the RecSys candidates are modeled as a stochastic variable $\boldsymbol \xi = \boldsymbol \mu + \boldsymbol \epsilon$, where $\boldsymbol \mu \in \mathbb R^K$ is the mean of the process and $\boldsymbol \epsilon \sim \mathcal N(\boldsymbol 0, \Sigma)$ is a multi-variate Gaussian noise with covariance $\Sigma := [\sigma_{aa'}]$ and $\sigma_{aa'} := \operatorname{cov}(\epsilon_a, \epsilon_{a'})$. The key quantities are the discriminal dispersion $s_{aa'} := \sigma_{aa} + \sigma_{a'a'} - 2\sigma_{aa'}$ and the normalized score difference:

\begin{align}
\label{eq:normalized_z}
    z_{aa'} := \frac{\mu_a - \mu_{a'}}{\sqrt{s_{aa'}}}.
\end{align}
 
\paragraph{Outcome probabilities}
The outcome probabilities under the generalized Rao-Kupper model are:
\begin{align*}
    &P_{\boldsymbol \theta}(H_t\!=\!1 | A_t\!=\!(a,a')) = \frac{1}{1 + e^{-(z_{a'a} - \eta_{a'a})}}
    \\[3pt]
    &P_{\boldsymbol \theta}(H_t\!=\!0 | A_t\!=\!(a,a')) = \frac{1}{1 + e^{-(z_{aa'} - \eta_{aa'})}} \\
    &P_{\boldsymbol \theta}(H_t\!=\!0.5 | A_t\!=\!(a,a')) = \\
    &\qquad \frac{e^{2\eta_{aa'}} - 1}{\big(1 + e^{-(z_{aa'} - \eta_{aa'})}\big)\big(1 + e^{-(z_{a'a} - \eta_{a'a})}\big)},
\end{align*}
where $\eta_{aa'} = \eta_{a'a} \geq 0$ are pair-specific tie thresholds. Intuitively, when the normalized score difference $z_{aa'}$ falls within the threshold $\eta_{aa'}$, the evaluator cannot distinguish the RecSys candidates, producing a tie.
 
\paragraph{Low-rank parameterization}
To control the complexity of the model, both tie thresholds and covariance are parameterized via low-rank factor models. Let $\boldsymbol G \in \mathbb{R}^{C \times k_{\text{tie}}}$ denote the matrix of parameters to estimate and $\boldsymbol \Phi \in \mathbb{R}^{C \times k_{\text{tie}}}$ be a constant matrix of rank $k_{\text{tie}} \leq A$ where the elements are given by a discrete cosine transform of the fourth type basis:
\begin{align*}
    \phi_{aa'} = \sqrt{\frac{2}{A}} \cos \left( \frac{\pi}{4A}(2a-1)(2a'-1) \right)\\
    \qquad \qquad \forall a, a'\in [C].
\end{align*}
The tie thresholds are constructed as:
\begin{align*}
    \eta_{aa'} = \boldsymbol{g}_a \cdot \boldsymbol{\phi}_{a'} + \boldsymbol{g}_{a'} \cdot \boldsymbol{\phi}_a,
\end{align*}
where $\mathbf{g}_a$ and $\boldsymbol{\phi}_a$ respectively denote the $a$-th line of matrices $\boldsymbol G$ and $\boldsymbol \Phi$. The covariance is parameterized as $\Sigma = \boldsymbol D + \boldsymbol \Lambda \boldsymbol \Lambda^\top$, where $\boldsymbol D$ is a positive diagonal matrix and $\boldsymbol \Lambda \in \mathbb{R}^{A \times k_{\text{cov}}}$. The ranks $k_{\text{tie}}$ and $k_{\text{cov}}$ control complexity: $k_{\text{tie}} = 0$ recovers the original single-threshold model, while $k_{\text{cov}} = 0$ yields independent RecSys candidates.
 
\paragraph{Estimation and ranking}
The resulting model parameters $\hat{\boldsymbol \theta} = (\hat{\boldsymbol \mu}, \boldsymbol G, \boldsymbol \Phi, \boldsymbol \Lambda)$ are estimated by maximizing the logarithm of the likelihood (Eq.~\ref{eq:likelihood}). The estimated candidate strengths are $\hat{\boldsymbol \xi} = \hat{\boldsymbol \mu}$, consistent with the generic notation of Section~\ref{sec:ranking}. Let $\hat{V}$ denote the sandwich variance estimator~\citep{huber1967behavior} of $\hat{\boldsymbol{\xi}}$. For each RecSys $a$, we construct marginal confidence intervals $\mathcal{C}_a := [\hat{\xi}_a \pm z_{1-\alpha/2} \sqrt{\hat{V}_{aa}}]$. The approximate rank of RecSys $a$ is:
\begin{align}
\label{eq:approx_rank}
    \operatorname{rank}(\hat{\boldsymbol{\xi}})_a := 1 + \sum_{\mathclap{a' \in [C]}} \mathbbm{1}\{ \inf \mathcal{C}_{a'} > \sup \mathcal{C}_a \},
\end{align}
such that the best RecSys is ranked 1 and candidates with overlapping confidence intervals share the same rank.

\section{Tasks and Datasets}
\label{app:datasets}
 
We formalize the main slate recommendation task evaluated in the body of the paper (\emph{Joint Selection and Ordering}) and the two auxiliary slate problems (\emph{Set Selection}, \emph{Reranking}) that we report on for robustness in Appendix~\ref{app:auxiliary-tasks}. Table~\ref{tab:datasets} summarizes their domains and statistics.

\paragraph{Task-difficulty levels}

\begin{figure}[h]
  \centering
  \includegraphics[width=0.8\linewidth]{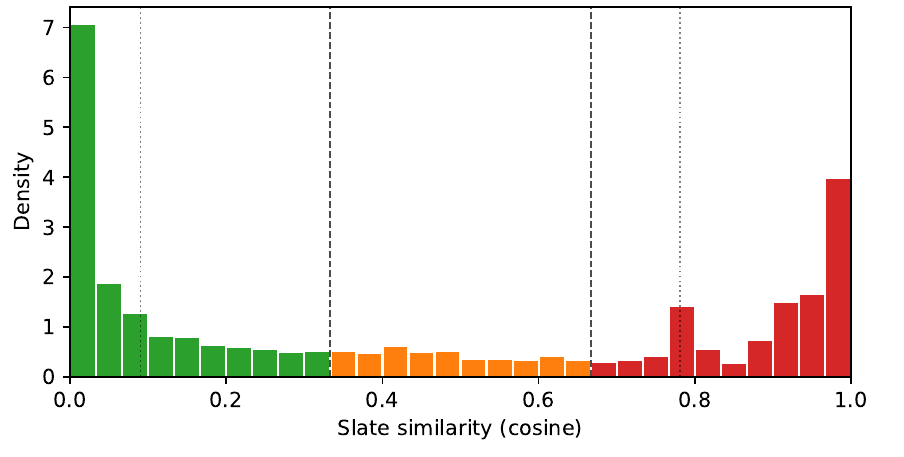}
  \caption{Distribution of similarities between all pairs of slates in the two datasets of the main task (Section~\ref{app:task-joint}).} 
  \label{fig:similarity_distribution}
\end{figure}

We discretize a slate-similarity proxy $\operatorname{sim}: \mathcal L \times \mathcal L \to [0, 1]$ into slate-comparison difficulty:
\begin{align*}
\operatorname{diff}(L, L') := \begin{cases}\text{low} & \operatorname{sim}(L, L') \in [0, 1/3) \\ \text{med.} & \operatorname{sim}(L, L') \in [1/3, 2/3) \\ \text{high} & \operatorname{sim}(L, L') \in [2/3, 1].\end{cases}
\end{align*}
The thresholds isolate the two modes of the empirical similarity distribution in the main task (Figure~\ref{fig:similarity_distribution}).

\subsection{Joint Selection and Ordering (main task)}
\label{app:task-joint}
A realistic recommender must simultaneously select a subset of items and arrange them. The agent outputs an ordered list
\(L_u = \langle i_1, \dots, i_k \rangle \subseteq \mathcal{I}\),
combining both \emph{what} to recommend and \emph{how} to order it. This setting mirrors practical scenarios such as movie carousels, news feeds, or playlist generation—where both composition and order drive user satisfaction. We evaluate on \textbf{MIND}~\cite{wuMINDLargescaleDataset2020} (click chronologies for news articles) and \textbf{Spotify Million}~\cite{chenRecsysChallenge20182018} (curator-defined playlist orders). Ground truth is computed using the nDCG of the proposed list against the reference order.
 
\subsection{Set Selection (auxiliary)}
\label{app:task-set}
Given the entire catalogue \(\mathcal{I}\) and a small budget \(k \ll |\mathcal{I}|\),
the recommender must return an unordered set
\(S_u \subset \mathcal{I},\; |S_u| = k\),
that maximizes user-specific utility. In this setting, the model decides \emph{what} to recommend but not \emph{how} to order it. We evaluate on two canonical rating datasets: \textbf{MovieLens 1M}~\cite{harperMovieLensDatasetsHistory2015} and \textbf{Amazon23}~\cite{houBridgingLanguageItems2024}. Ground-truth utility is defined as the sum of rescaled item ratings.
 
\subsection{Reranking (auxiliary)}
\label{app:task-rerank}
Here the item set is fixed in advance (e.g., a playlist draft or news slate); the model decides only the \emph{permutation}
\(\pi_u : \{1, \dots, k\} \to S_u\)
that maximizes downstream engagement. We adopt the same two datasets used for Joint Selection and Ordering—\textbf{MIND} and \textbf{Spotify}—since they provide an authoritative reference order. Ground truth is the nDCG of the proposed permutation against the reference.
 
\begin{table}[h]
\centering
\small
\setlength{\tabcolsep}{5pt}
\caption{Validation datasets. \emph{Joint} = Joint Selection and Ordering (main task); \emph{Set} and \emph{Rerank} are auxiliary tasks reported in Appendix~\ref{app:auxiliary-tasks}.}
\label{tab:datasets}
\resizebox{\columnwidth}{!}{%
\begin{tabular}{lccccc}
\toprule
\textbf{Task} & \textbf{Domain} & \textbf{Dataset} & \textbf{Users} & \textbf{Items} & \textbf{Interactions} \\
\midrule
Joint  & News       & MIND               & 50k & 65k & Click / Order \\
Joint  & Music      & Spotify Pod 20k    & 19k & 20k & Playlist order \\
Set    & Movies     & MovieLens-1M       & 6k  & 3.4k & Rating (1–5) \\
Set    & E-commerce & Amazon23           & 51k & 23k & Rating (1–5) \\
Rerank & News       & MIND               & 50k & 65k & Click / Order \\
Rerank & Music      & Spotify Pod 20k    & 19k & 20k & Playlist order \\
\bottomrule
\end{tabular}}
\end{table}

\section{Evaluation protocol: slate construction, pairwise queries, and leakage mitigation}
\label{app:eval-protocol}

\paragraph{User sampling and train/test split.}
For every (task, dataset) condition we sample $N = 100$ users uniformly without replacement from the catalogue with a fixed seed. Each user's interaction history is split into a training portion (used to fit the candidate RecSys) and a held-out test portion (used to compute the ground-truth empirical utility, Def.~\ref{def:empirical_utility}).

\paragraph{Candidate slate construction.}
For each (user, RecSys) pair we materialise a single ranked slate of length $K = 10$ (with $K \in \{2, 5\}$ in the slate-size sensitivity study, App.~\ref{app:auxiliary-tasks}). The slate is produced by the RecSys' \texttt{recommend\_topk} routine over the full catalogue, with the user's training items excluded from the candidate pool so that the LLM never re-judges an item the model has already seen during training. All  candidate RecSys consume the \emph{same} train/test split and the \emph{same} exclusion mask, ensuring an equal comparison.

\paragraph{Pairwise comparison budget per user.}
With $C = 6$ candidate RecSys, each user triggers $\binom{C}{2} = 15$ unordered pairs of slates. Each pair is judged in both orientations (LL' and L'L) to cancel positional bias (Section \ref{sec:llm-ensemble}), and each orientation aggregates $M$ independent ensemble votes (default $M = 6$). This yields $\binom{C}{2} \cdot 2 \cdot M = 180$ LLM forward passes per user, and $N \cdot \binom{C}{2} \cdot 2 \cdot M = 18{,}000$ forward passes per experimental condition. 

\paragraph{Infrastructure}
All experiments were run on NVIDIA H100 GPUs. LLM inference uses the vLLM  backend \citep{kwonEfficient2023} with FlashAttention-2 \citep{daoFlashAttention22024}; RecSys training and offline-policy estimators (IPS, PI) run on CPU-only with Intel Xeon Gold 6448Y.

\paragraph{What the LLM actually sees.}
The slates handed to the LLM contain only \emph{natural-language metadata} extracted from each dataset's item table: news headlines for \textsc{MIND}; track titles and artist names (and, when available, album names) for \textsc{Spotify}; movie titles and genre tags for \textsc{MovieLens1M}; product titles for \textsc{Amazon23}. Item identifiers are never exposed in the prompt. The user context block is rendered from the training-portion interactions as a \emph{history} of (item-metadata, implicit-feedback) pairs, where implicit feedback is binarised to $0$ (skipped/disliked) or $1$ (clicked/loved). All textual fields are passed verbatim --- we do not summarise or paraphrase --- so that the LLM judge operates on the same surface representation it would encounter in a production semantic-RecSys interface.

\paragraph{Sampling protocol for the internal coherence axioms.}
\label{app:axiom-sampling}
The internal coherence axioms (Eq. \ref{def:irreflexivity_score}, \ref{def:asymmetry_score}, \ref{def:transitivity_score}) iterate over slate pairs (asymmetry) and triples (transitivity) over $\bar{\mathcal L}_n$, whose cardinality scales as $\mathcal O(|\bar{\mathcal L}_n|)$, $\mathcal O(|\bar{\mathcal L}_n|^2)$ and $\mathcal O(|\bar{\mathcal L}_n|^3)$ respectively. With $N{=}100$ users and a single representative slate per user, the asymmetry domain reaches $\binom{N}{2}{\approx}5{,}000$ and the transitivity domain $\binom{N}{3}{\approx}1.6{\times}10^{5}$, which is intractable to evaluate exhaustively at $M{=}6$ LLM votes per comparison. We adopt the following per-axiom protocol:
\begin{itemize}
\setlength{\itemsep}{0pt}
\setlength{\parskip}{0pt}
  \item \textbf{Irreflexivity} (Def.~\ref{def:irreflexivity_score}): evaluated \emph{exhaustively} (one self-comparison per user, $N$ comparisons total). The domain is $\mathcal O(N)$ and requires no sampling.
  \item \textbf{Asymmetry} (Def.~\ref{def:asymmetry_score}): $M_{\text{asym}} = 500$ pairs per condition. The sample is the union of an \emph{essential chain} $\{(L_n, L_{n+1})\}_{n=1}^{N-1}$ that guarantees each user appears in at least one tested pair, plus a uniform-random complement of $M_{\text{asym}} - (N-1)$ pairs. This stratification removes user-coverage bias from the asymmetry estimator.
  \item \textbf{Transitivity} (Def.~\ref{def:transitivity_score}): $M_{\text{trans}} = 100$ triples sampled uniformly without replacement from $\bar{\mathcal L}_n^{3}$.
\end{itemize}

\section{Votes}
\label{app:votes}

\begin{figure}[h]
  \centering
  \includegraphics[width=0.8\columnwidth]{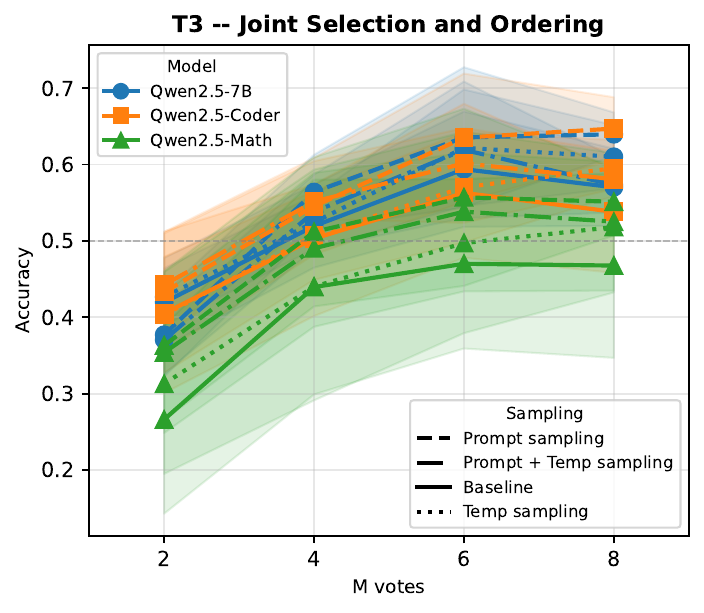}
  \caption{Pairwise accuracy vs.\ ensemble size $M$, by diversification strategy (vanilla, prompt, temperature, both).}
  \label{fig:vote_accuracy_by_task}
\end{figure}

Figure~\ref{fig:vote_accuracy_by_task} shows the impact of the ensemble size $M$ on the external accuracy (Def.~\ref{def:ext_accuracy}) across our four diversification strategies (vanilla, prompt, temperature, both). Across every dataset and model, accuracy increases sharply between $M{=}1$ and $M{=}6$, then flattens and effectively plateaus from $M{\approx}6$ onward.

This shape is consistent with the scaling-law analysis of compound inference systems developed by~\citet{chen2024are}. They show that majority-vote ensembles do not exhibit a monotone ``more calls is better'' behavior: the accuracy of the aggregated verdict is governed by a Bernoulli concentration around the underlying per-call success rate, so additional votes mostly reduce the variance of the aggregation rather than shift its mean. Once enough votes have been drawn to push the majority outcome close to that limiting rate, further calls leave the verdict essentially unchanged. The aggregate curve over a heterogeneous workload is therefore predicted to rise steeply at small $M$ and then flatten -- which is exactly the regime we observe, with the knee located near $M{\approx}6$.

This analysis motivates our default choice of $M{=}6$ throughout the paper: it sits at the knee of the curve, capturing essentially all of the variance reduction available from ensembling while keeping the per-comparison cost bounded ($2M$ forward passes per slate pair, see App.~\ref{app:eval-protocol}). Pushing $M$ further would multiply compute without measurably improving accuracy.

\paragraph{Measured cost and comparison to a live A/B test.}

\section{Auxiliary Tasks Results}
\label{app:auxiliary-tasks}
 
To corroborate that the conclusions drawn on the main \emph{Joint Selection and Ordering} task are not artifacts of that specific setting, we report the same external coherence diagnostics on two auxiliary slate problems: \emph{Set Selection} (Appendix~\ref{app:task-set}) on \textsc{Amazon23} and \textsc{MovieLens1m}, and \emph{Reranking} (Appendix~\ref{app:task-rerank}) on \textsc{MIND} and \textsc{Spotify}. The full Spearman correlation table including the main task and both auxiliary tasks is reported in Table~\ref{tab:spearman_external_full}.
 
\paragraph{Quality gap and judgment accuracy across all three tasks}
The negative correlation $\rho(\text{gap}, \text{acc})$ observed in the body extends to both auxiliary tasks: across all 18 conditions (3 tasks $\times$ 2 datasets $\times$ 3 models), the correlation is significant ($p < 0.001$) and ranges from $-0.36$ to $-0.67$. Two qualitative trends emerge. First, $|\rho(\text{gap}, \text{acc})|$ is largest on tasks with stronger ordering signal (Reranking and Joint Selection and Ordering), suggesting that the LLM-Ensemble extracts more calibrated signal when positional structure is available. 
 
\begin{figure}[h]
  \centering
  \includegraphics[width=0.8\linewidth]{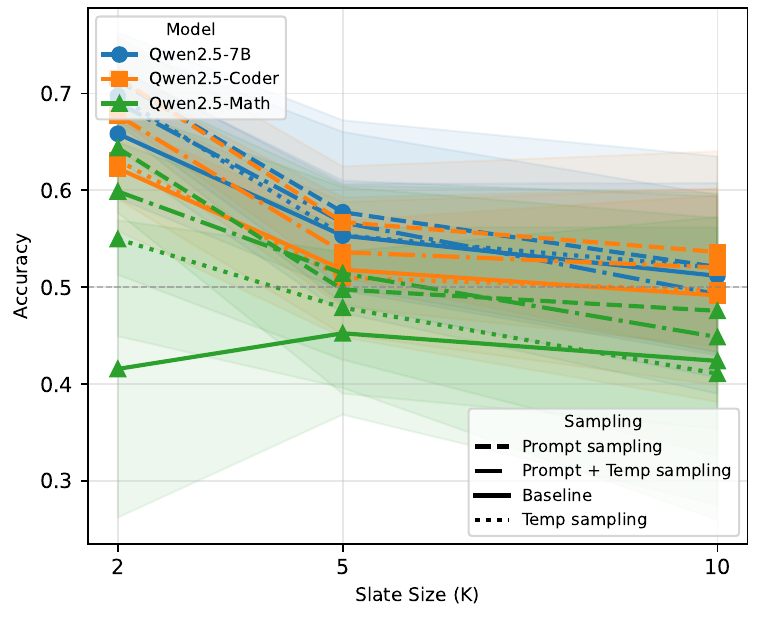}
  \caption{Pairwise accuracy vs.\ slate size $K$. Diversification keeps its margin over vanilla at every $K$.}
  \label{fig:accuracy_by_slate_size}
\end{figure}

\paragraph{Effect of slate size on Set Selection and Reranking}
The accuracy degradation with slate size $K$ reported in the body for Joint Selection and Ordering (Figure~\ref{fig:accuracy_by_slate_size}) is most pronounced on Set Selection, where no positional anchor compensates for the additional combinatorial load. Reranking, which preserves the full positional structure of the comparison, exhibits a smaller drop than Joint Selection.

\paragraph{Confidence-difficulty interaction across tasks}
On Reranking, low-difficulty/high-confidence comparisons reach 0.89 accuracy while high-difficulty/low-confidence drop to 0.59; high-confidence judgments on hard pairs still reach 0.77 (similar pattern to Joint Selection). On Set Selection, accuracy stays modest across all confidence$\times$difficulty cells (0.15--0.64), consistent with the intrinsic difficulty of unordered set comparison at small $K$. These results reinforce the body's claim that vote margin can be used as a reliability indicator, with a stronger discriminative effect on tasks that include ordering.

\begin{figure*}[h]
  \centering
  \includegraphics[width=\textwidth]{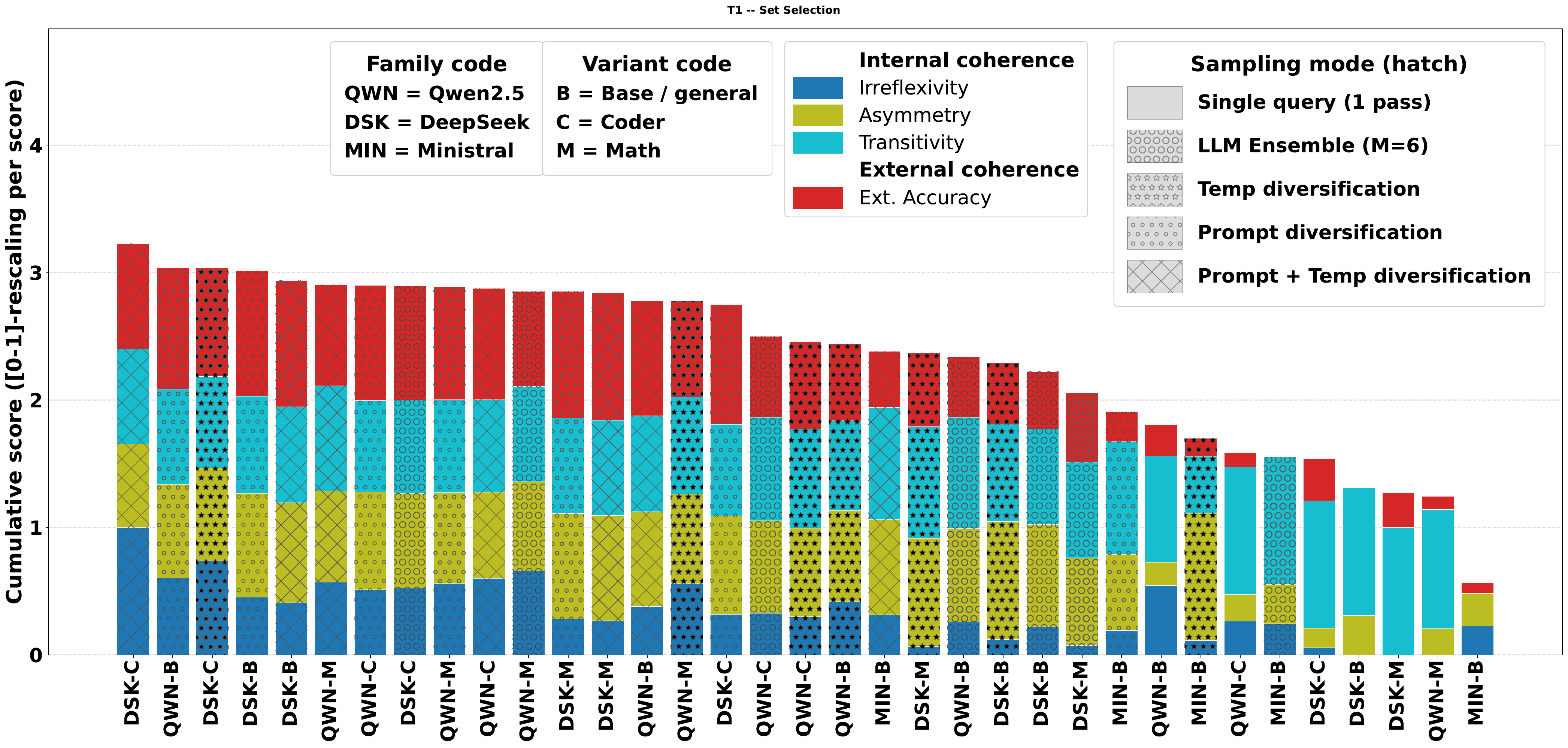}
  \caption{External (accuracy) and internal coherence (irreflexivity, asymmetry, transitivity) for all configurations on Task 1.}
  \label{fig:axiom_stacked_bars_wuniform_task1}
\end{figure*}

\begin{figure*}[h]
  \centering
  \includegraphics[width=\textwidth]{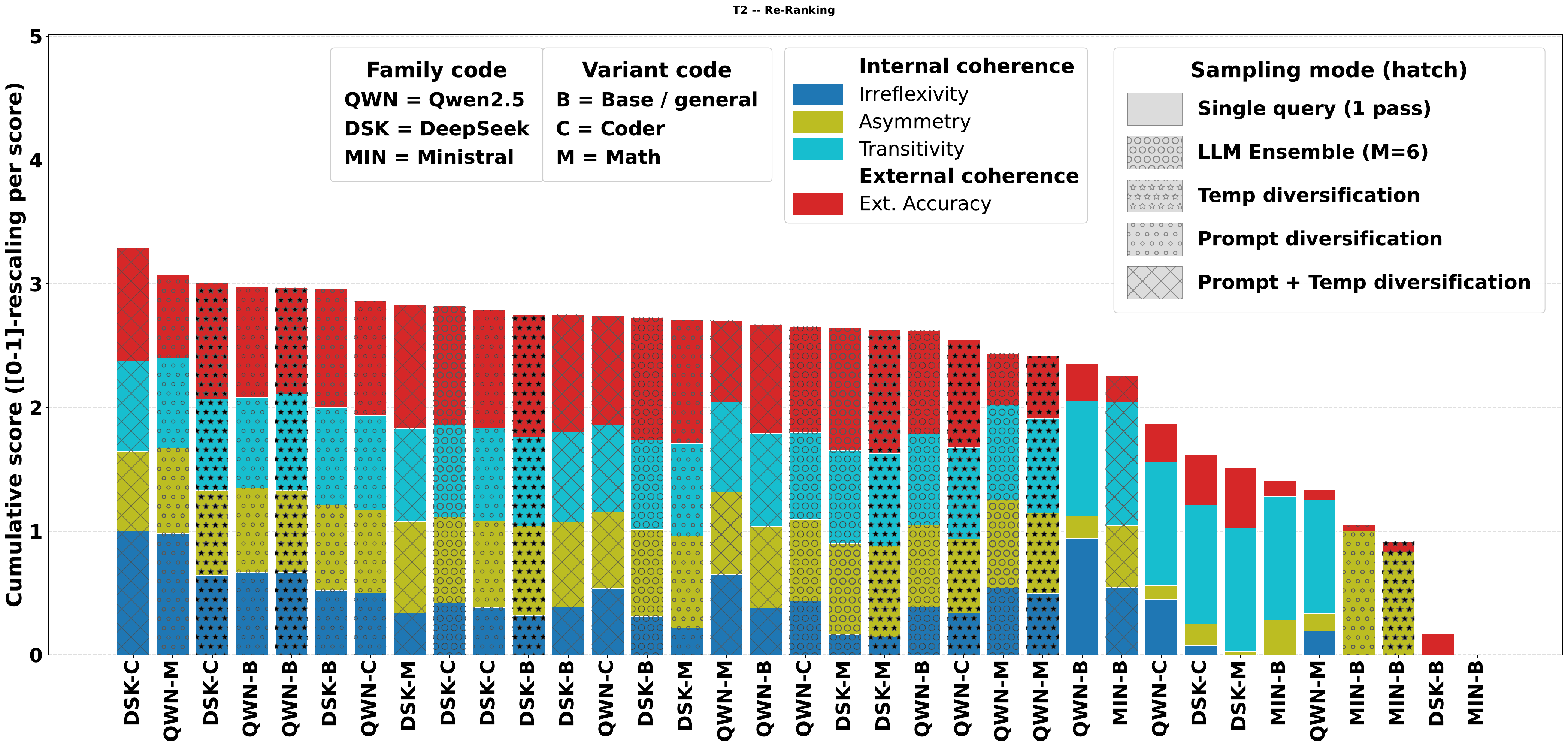}
  \caption{External (accuracy) and internal coherence (irreflexivity, asymmetry, transitivity) for all configurations on Task 2.}
  \label{fig:axiom_stacked_bars_wuniform_task2}
\end{figure*}

\begin{table}[t]
\centering
\small
\caption{Spearman correlations between absolute utility gaps and the ensemble correctness ($\rho_{\operatorname{Acc}}$).
Significant ($p < 0.05$) values in \textbf{bold}.}
\label{tab:spearman_external}
\resizebox{0.8\columnwidth}{!}{
\begin{tabular}{llcc}
\toprule
\textbf{Dataset} & \textbf{Model}
& $\rho_{\operatorname{Acc}}$ & $p$ \\
\midrule
\multirow{7}{*}{MIND}
& Qwen2.5-7B      & \textbf{-0.508} & $<.001$ \\
& Qwen2.5-Coder   & \textbf{-0.502} & $<.001$ \\
& Qwen2.5-Math    & \textbf{-0.553} & $<.001$ \\
& DeepSeek-LLM    & \textbf{-0.543} & $<.001$ \\
& DeepSeek-Coder  & \textbf{-0.563} & $<.001$ \\
& DeepSeek-Math   & \textbf{-0.520} & $<.001$ \\
& Ministral-8B    & \textbf{-0.653} & $<.001$ \\
\cmidrule(lr){1-4}
\multirow{7}{*}{Spotify}
& Qwen2.5-7B      & \textbf{-0.542} & $<.001$ \\
& Qwen2.5-Coder   & \textbf{-0.549} & $<.001$ \\
& Qwen2.5-Math    & \textbf{-0.549} & $<.001$ \\
& DeepSeek-LLM    & \textbf{-0.552} & $<.001$ \\
& DeepSeek-Coder  & \textbf{-0.614} & $<.001$ \\
& DeepSeek-Math   & \textbf{-0.538} & $<.001$ \\
& Ministral-8B    & \textbf{-0.685} & $<.001$ \\
\bottomrule
\end{tabular}}
\end{table}

\begin{table}[t]
\centering
\small
\caption{Spearman $\rho$ between quality gap and LLM signals across two auxiliary tasks (pooled across vote counts and slate sizes). $\rho(\text{gap}, \text{conf})$: correlation between ground-truth quality gap and vote confidence; $\rho(\text{gap}, \text{acc})$: correlation between quality gap and judgment accuracy. Statistically significant correlations ($p < 0.05$) are \textbf{bolded}.}
\label{tab:spearman_external_full}

\resizebox{0.9\columnwidth}{!}{
\begin{tabular}{lllrr}
\toprule
\textbf{Task} & \textbf{Dataset} & \textbf{Model}
& $\rho(\text{gap,acc})$ & $p$ \\
\midrule

\multirow{14}{*}{\shortstack[l]{Set\\Selection}}
& \multirow{7}{*}{Amazon23}
& Qwen2.5-7B      & \textbf{-0.465} & $<.001$ \\
&                 & Qwen2.5-Coder  & \textbf{-0.455} & $<.001$ \\
&                 & Qwen2.5-Math   & \textbf{-0.476} & $<.001$ \\
&                 & DeepSeek-Coder & \textbf{-0.510} & $<.001$ \\
&                 & DeepSeek-LLM   & \textbf{-0.463} & $<.001$ \\
&                 & DeepSeek-Math  & \textbf{-0.446} & $<.001$ \\
&                 & Ministral-8B   & \textbf{-0.577} & $<.001$ \\

\cmidrule(lr){2-5}

& \multirow{7}{*}{MovieLens}
& Qwen2.5-7B      & \textbf{-0.501} & $<.001$ \\
&                 & Qwen2.5-Coder  & \textbf{-0.451} & $<.001$ \\
&                 & Qwen2.5-Math   & \textbf{-0.518} & $<.001$ \\
&                 & DeepSeek-Coder & \textbf{-0.504} & $<.001$ \\
&                 & DeepSeek-LLM   & \textbf{-0.537} & $<.001$ \\
&                 & DeepSeek-Math  & \textbf{-0.523} & $<.001$ \\
&                 & Ministral-8B   & \textbf{-0.504} & $<.001$ \\

\midrule

\multirow{14}{*}{Reranking}
& \multirow{7}{*}{MIND}
& Qwen2.5-7B      & \textbf{-0.585} & $<.001$ \\
&                 & Qwen2.5-Coder  & \textbf{-0.545} & $<.001$ \\
&                 & Qwen2.5-Math   & \textbf{-0.715} & $<.001$ \\
&                 & DeepSeek-Coder & \textbf{-0.614} & $<.001$ \\
&                 & DeepSeek-LLM   & \textbf{-0.615} & $<.001$ \\
&                 & DeepSeek-Math  & \textbf{-0.555} & $<.001$ \\
&                 & Ministral-8B   & \textbf{-0.704} & $<.001$ \\

\cmidrule(lr){2-5}

& \multirow{7}{*}{Spotify}
& Qwen2.5-7B      & \textbf{-0.612} & $<.001$ \\
&                 & Qwen2.5-Coder  & \textbf{-0.594} & $<.001$ \\
&                 & Qwen2.5-Math   & \textbf{-0.658} & $<.001$ \\
&                 & DeepSeek-Coder & \textbf{-0.635} & $<.001$ \\
&                 & DeepSeek-LLM   & \textbf{-0.603} & $<.001$ \\
&                 & DeepSeek-Math  & \textbf{-0.579} & $<.001$ \\
&                 & Ministral-8B   & \textbf{-0.723} & $<.001$ \\

\bottomrule
\end{tabular}}
\end{table}
 
\section{Composite-Utility Weight Regimes: Sensitivity Analysis}
\label{app:weight_regimes}

The composite empirical utility (Def.~\ref{def:empirical_utility}) requires fixing a scalarization weight vector $\boldsymbol \alpha = (\alpha_{\text{nDCG}}, \alpha_{\text{Diversity}}, \alpha_{\text{Novelty}}, \alpha_{\text{Serendipity}})$. We report all main-paper results under an \emph{nDCG-focused} weighting because relevance is the dominant first-order objective in deployed RecSys pipelines ~\citep{}; the three remaining axes act as complementary, smaller-weight refinements. To assess the robustness of our conclusions to this modeling choice, we additionally re-run the full benchmark under two others contrasting regimes:

\begin{table}[t]
\centering
\small
\setlength{\tabcolsep}{4pt}

\begin{tabular}{lcccc}
\toprule
\textbf{Regime} & $\alpha_{\text{nDCG}}$ & $\alpha_{\text{Diversity}}$ & $\alpha_{\text{Novelty}}$ & $\alpha_{\text{Serendipity}}$ \\
\midrule
\emph{accuracy} (main) & 0.70 & 0.10 & 0.10 & 0.10 \\
\emph{uniform}         & 0.25 & 0.25 & 0.25 & 0.25 \\
\emph{diversity}       & 0.10 & 0.40 & 0.25 & 0.25 \\
\bottomrule
\end{tabular}

\caption{Reward weight configurations for the different evaluation regimes.}
\label{tab:reward_weights}
\end{table}

\begin{figure*}[h]
  \centering
  \includegraphics[width=\textwidth]{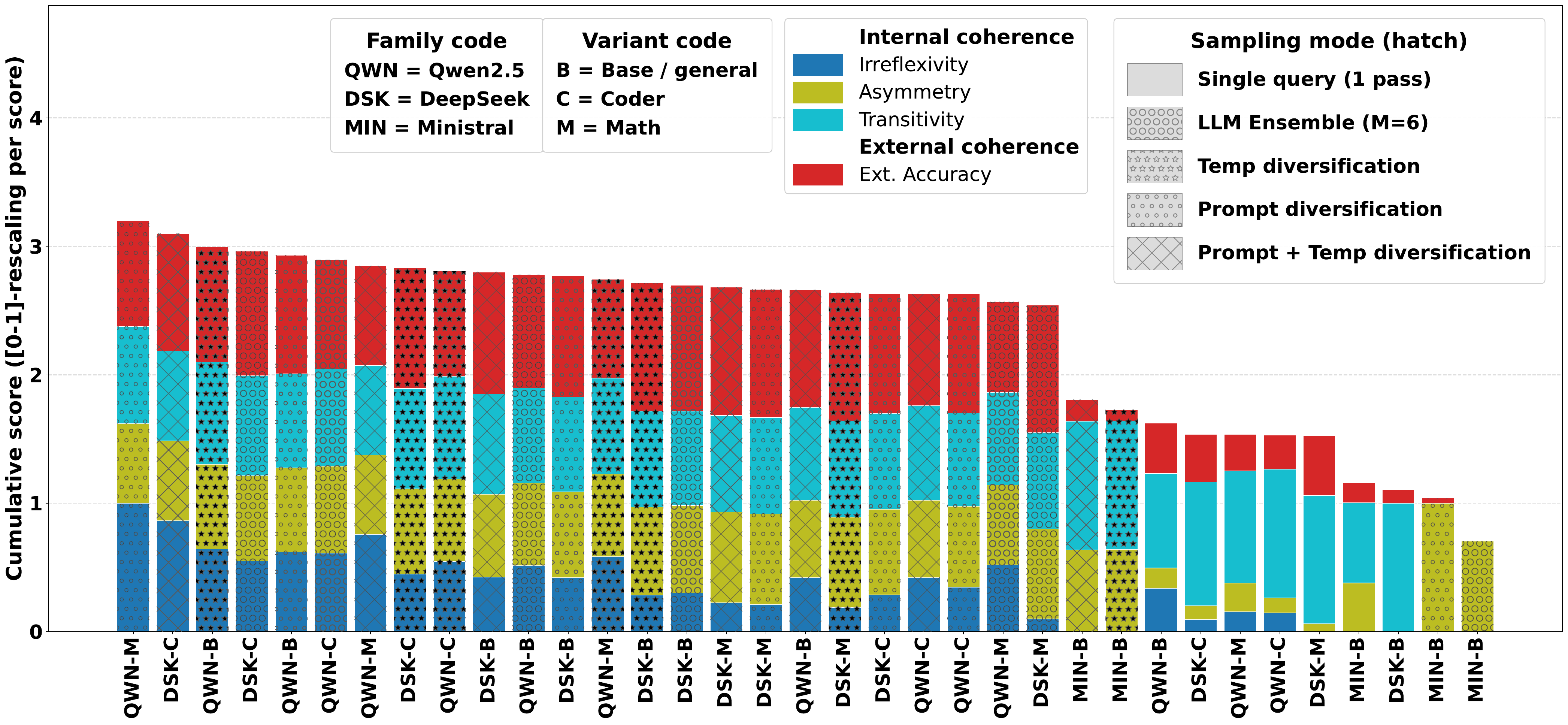}
  \caption{External (accuracy) and internal coherence (irreflexivity, asymmetry, transitivity) for all configurations. With a uniform weigth distribution.}
  \label{fig:axiom_stacked_bars_wuniform}
\end{figure*}

\begin{figure*}[h]
  \centering
  \includegraphics[width=\textwidth]{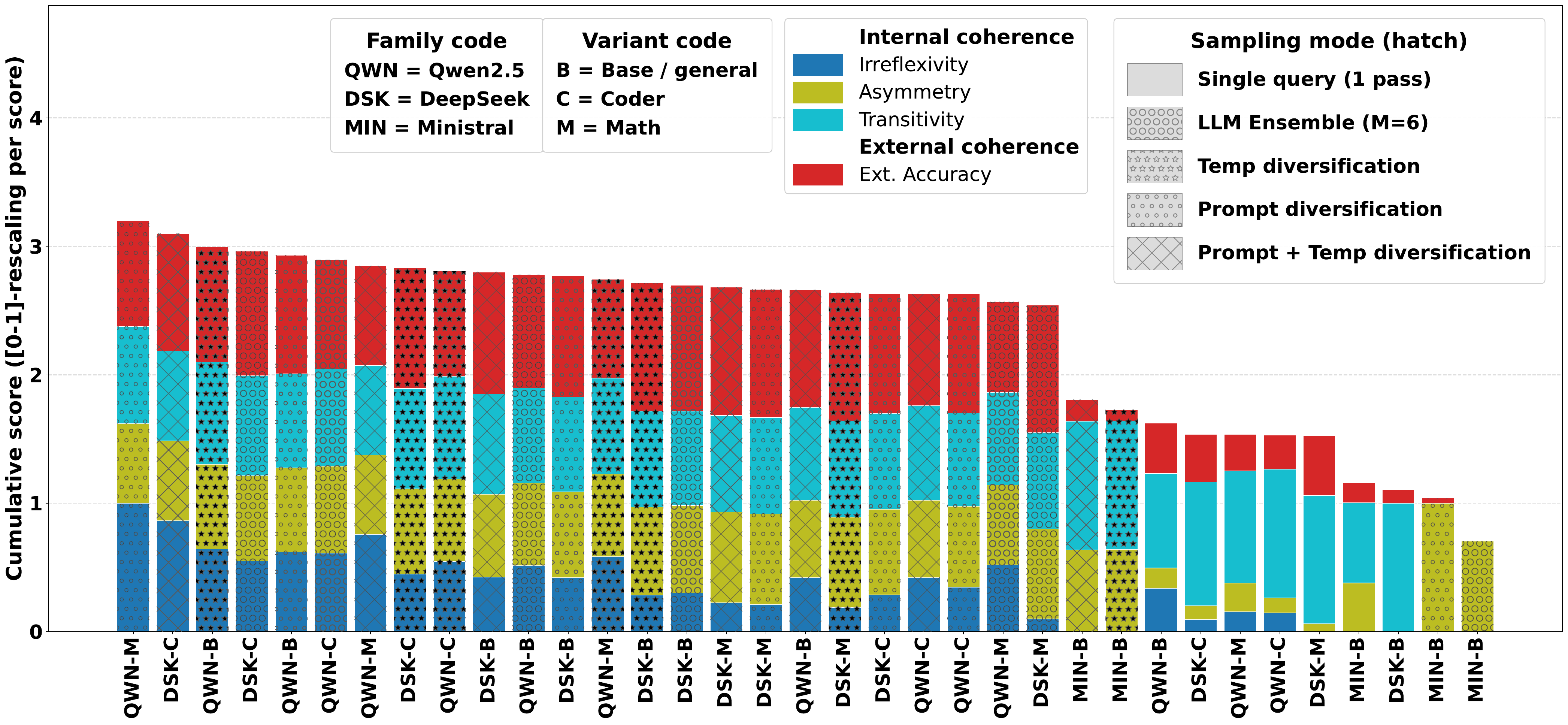}
  \caption{External (accuracy) and internal coherence (irreflexivity, asymmetry, transitivity) for all configurations. With a diversity weigth distribution.}
  \label{fig:axiom_stacked_bars_wdiversity}
\end{figure*}

\paragraph{Axiom contributions per regime.}

Figures~\ref{fig:axiom_stacked_bars_waccuracy}, \ref{fig:axiom_stacked_bars_wuniform}, and~\ref{fig:axiom_stacked_bars_wdiversity} display the per-axiom contributions to the scalarisation weight for each (model $\times$ sampling mode) configuration under the three regimes (slate size $K{=}10$, ensemble size $M{=}6$). The qualitative ordering of configurations is preserved across regimes: high-performing configurations remain dominant whether nDCG or Diversity is up-weighted.

 \section{Prompt Templates and Sampling Mechanisms}
\label{app:prompts}

This appendix presents the prompt templates used across all benchmarking tasks
and datasets, the eight reasoning-approach variants, and the
temperature-sampling regime used to study verdict stability. Templates are
presented in their \emph{maximally general form} so practitioners can reuse
them directly: domain-specific terms appear as
{\color{placeholdercolor}\ttfamily blue placeholders}, and chat-template tokens
appear as {\color{tagcolor}\ttfamily red placeholders} that the reader
substitutes for their target model's conventions (e.g.\ Qwen's
\icode{<|im\_start|>}/\icode{<|im\_end|>}, Llama-3's
\icode{<|start\_header\_id|>}/\icode{<|eot\_id|>}, ChatML, etc.).


\paragraph{Prompt design principles}

Each evaluation call follows a generic two-turn chat structure, agnostic of
both the application domain and the underlying chat-model family:

\begin{enumerate}[leftmargin=*, itemsep=2pt]
  \item \textbf{System turn} — establishes the LLM's domain persona
        (\placeholder{domain\_expert\_role}) and bounds evaluation to
        \emph{verifiable} knowledge of \placeholder{domain}.
  \item \textbf{User turn} — supplies (i)~the user's interaction history with
        an explicit \placeholder{signal\_type} (e.g.\ explicit ratings,
        implicit clicks, ranked orders, dwell time), (ii)~the candidate
        recommendation lists, (iii)~a four-step evaluation rubric, and
        (iv)~the required output format.
\end{enumerate}
Figure~\ref{fig:prompt:system} shows the generic prompt used in the system turn. User turn prompts are described in the following sections.

The four dimensions of the evaluation rubric — \emph{Alignment}, \emph{Diversity},
\emph{Quality}, \emph{Authenticity} — are domain-independent and applicable to any
ranked-item recommendation task.They are intentionally chosen to mirror the multi-criterion rubric paradigm popularised by LLM-as-judge frameworks such as
G-Eval~\citep{liuGEval2023}, while binding each criterion to a well-established recommender-systems quality axis: \emph{Alignment} corresponds to slate-level \emph{relevance} to the user history (the nDCG component of the composite utility, App.~\ref{app:external-metrics}); \emph{Diversity} covers intra-list dissimilarity, novelty and serendipity, which form the canonical
``beyond-accuracy'' axes of recommender evaluation~\citep{kaminskas2017diversity, CastellsNovelty22}; \emph{Quality} reflects per-item production/critical signals frequently used as exogenous proxies in deployed pipelines; and \emph{Authenticity} guards against the item-fabrication / out-of-catalogue hallucinations that have been identified as a first-order failure mode of LLM-based recommenders~\citep{JiangBeyongUtility2025}.

\subsection{Pairwise Evaluation Prompt (Generic Template)}
\label{app:prompt-pair}

Figure~\ref{fig:prompt:template} shows the canonical pairwise prompt. All blue bracketed tokens 
are runtime placeholders; all red bracketed tokens are chat-template markers
to be substituted for the target model.

\begin{figure*}
\begin{systemprompt}[System Turn — Generic]
{\color{tagcolor}<SYS\_BEGIN>}\\
You are an expert \placeholder{domain} recommendation evaluator with deep
knowledge of \placeholder{domain\_knowledge\_scope} (e.g.\ history, categories,
creators, and user preferences).{\color{tagcolor}<SYS\_END>}
\end{systemprompt}
\caption{Template prompt for the system turn.}
\label{fig:prompt:system}
\end{figure*}

\vspace{0.3em}

\begin{figure*}
\begin{userprompt}[User Turn — Pairwise Comparison (Generic)]
{\color{tagcolor}<USER\_BEGIN>}\\
{\color{tagcolor}<TASK>}\\
Compare two \placeholder{item\_type} recommendation lists based on verifiable
\placeholder{domain} knowledge and user satisfaction potential.\\
{\color{tagcolor}</TASK>}\\
\mbox{}\\
{\color{tagcolor}<USER\_PROFILE>}\\
\textbf{User \placeholder{interaction\_verb} History}
(Most Recent First)\\
\placeholder{signal\_description}
\ \emph{(e.g.\ ``Satisfaction scores: 1 (strongly disliked) to 5 (highly
appreciated)'', ``Click signal: 0/1'', ``Ranked order: 1 = preferred'')}\\
\mbox{}\\
\placeholder{history}\\
{\color{tagcolor}</USER\_PROFILE>}\\
\mbox{}\\
{\color{tagcolor}<CANDIDATE\_LISTS>}\\
**List 1:**\\
\placeholder{list1}\\
\mbox{}\\
**List 2:**\\
\placeholder{list2}\\
{\color{tagcolor}</CANDIDATE\_LISTS>}\\
\mbox{}\\
{\color{tagcolor}<EVALUATION\_FRAMEWORK>}\\
\textbf{Step 1: Item Authenticity Check}\\
- Identify any titles that appear fabricated, nonsensical, or unverifiable\\
- Flag any suspicious metadata (\placeholder{metadata\_fields},
e.g.\ year, creator, category combinations)\\
- Note: Fabricated items receive zero recommendation value\\
\mbox{}\\
\textbf{Step 2: User Preference Analysis}\\
Examine the user's history to infer:\\
- Preferred categories/\placeholder{taxonomy\_term} (informed by engagement signal)\\
- \placeholder{temporal\_or\_contextual\_dimension} preferences
\ \emph{(e.g.\ era, season, geography)}\\
- Quality expectations based on interaction patterns\\
- Creator/\placeholder{collection\_term} affinities
\ \emph{(e.g.\ franchise, artist, publisher, brand)}\\
\mbox{}\\
\textbf{Step 3: Multi-Objective List Quality Assessment}\\
Evaluate both lists on the four criteria:\\
\mbox{}\\
3.1 \textbf{Alignment} — Category/\placeholder{temporal\_or\_contextual\_dimension}
match, creator continuity, expected satisfaction given the user's history.\\
\mbox{}\\
3.2 \textbf{Diversity} — Variety, spread across
\placeholder{temporal\_or\_contextual\_dimension}, avoid near-duplicates
and redundant entries.\\
\mbox{}\\
3.3 \textbf{Quality} — \placeholder{quality\_signals\_for\_domain}
\ \emph{(e.g.\ critical acclaim, cultural impact, production pedigree,
chart performance, peer-review status)}.\\
\mbox{}\\
3.4 \textbf{Authenticity} — All items verifiable; fabricated = full penalty.\\
\mbox{}\\
\textbf{Step 4: Holistic Comparative Verdict}\\
Weigh all four dimensions and determine which list better serves this user.\\
{\color{tagcolor}</EVALUATION\_FRAMEWORK>}\\
\mbox{}\\
{\color{tagcolor}<OUTPUT\_FORMAT>}\\
First, write your step-by-step reasoning:\\
REASONING:\\
{[Your analysis. Cite specific items, reference user history. Min 3--4 sentences.]}\\
\mbox{}\\
Then, on a new line, your verdict:\\
VERDICT: 1 \ \ \textbf{or} \ \ VERDICT: 2\\
\mbox{}\\
Then, on a new line:\\
EXPLANATION: Brief explanation (40--50 words).\\
{\color{tagcolor}</OUTPUT\_FORMAT>}\\
\mbox{}\\
Begin your evaluation:{\color{tagcolor}<USER\_END>}\\
{\color{tagcolor}<ASSISTANT\_BEGIN>}
\end{userprompt}
\caption{Template prompt for a pairwise comparison between two slates for a given user.}
\label{fig:prompt:template}
\end{figure*}

\vspace{0.4em}
\paragraph{Placeholder dictionary}
Table~\ref{tab:placeholder-dict} maps each generic placeholder to the
instantiation used in our experiments, and lists alternatives for adjacent
domains. Practitioners adapting this template to a new dataset only need to
fill the dictionary; the rubric, output format, and chat structure remain
unchanged.

\begin{table*}[h]
\centering\small
\caption{Placeholder dictionary: generic prompt slots and instantiations across
domains. Only this table changes when porting the prompt to a new dataset.}
\begin{tabular}{@{}lll@{}}
\toprule
Placeholder & Our instantiation (MovieLens) & Alternative domains \\
\midrule
\placeholder{domain}              & film       & music, news, products, books \\
\placeholder{domain\_expert\_role}& film critic& music expert, news analyst, product evaluator \\
\placeholder{item\_type}          & movie      & track, article, product, paper \\
\placeholder{interaction\_verb}   & viewing    & listening, reading, purchase, citation \\
\placeholder{signal\_type}        & rating 1--5& click 0/1, rank order, play count \\
\placeholder{taxonomy\_term}      & genre      & genre, topic, category, field \\
\placeholder{temporal\_or\_contextual\_dimension} & era & period, region, season \\
\placeholder{collection\_term}    & franchise  & album, outlet, brand, lab \\
\placeholder{metadata\_fields}    & year, director, genre & artist, date, source, ISBN \\
\placeholder{quality\_signals\_for\_domain} & critical acclaim, production pedigree
                                  & chart rank, citations, peer review \\
\bottomrule
\end{tabular}
\label{tab:placeholder-dict}
\end{table*}

\paragraph{Chat-template marker substitutions}
The red markers above are replaced following
Table~\ref{tab:chat-template-markers}.

\begin{table}
\caption{Chat-template marker substitutions used to render the generic prompt into each model family's native format. For Mistral / Ministral, the markers collapse to \icode{[INST]} / \icode{[/INST]} pairs, with the system content prepended to the first user turn.}
\centering
\footnotesize
\setlength{\tabcolsep}{4pt}
\renewcommand{\arraystretch}{1.2}
\resizebox{\linewidth}{!}{%
\begin{tabular}{@{}l l@{}}
\toprule
Generic marker & Qwen 2.x / DeepSeek \\
\midrule
\icode{<SYS\_BEGIN>}  & \icode{<|im\_start|>system} \\
\icode{<SYS\_END>}    & \icode{<|im\_end|>} \\
\icode{<USER\_BEGIN>} & \icode{<|im\_start|>user} \\
\icode{<USER\_END>}   & \icode{<|im\_end|>} \\
\bottomrule
\end{tabular}}
\label{tab:chat-template-markers}
\end{table}
\subsection{Prompt Variants}
\label{app:prompt-variants}
 
To introduce diversity in the reasoning style while keeping the evaluation criteria
strictly constant, we maintain a pool of eight prompt variants. Each variant is
structurally identical to the generic prompt (Fig.~\ref{fig:prompt:template}): same
system turn, same user profile block, same candidate lists, same four-criterion
rubric, same output format. The \emph{sole} difference is a single instruction
placed inside the \texttt{<EVALUATION\_FRAMEWORK>} block, immediately before the
output format section. Figure~\ref{fig:prompt:variant_instructions} illustrates the block to add.
 
\vspace{0.4em}
\begin{figure*}
\begin{userprompt}[Specifying the reasoning approach in the user prompt]
\ldots\\
{\color{tagcolor}</EVALUATION\_FRAMEWORK>}\\
\mbox{}\\
\textit{REASONING APPROACH}\\
\mbox{}\\
\textit{$\langle$\,variant-specific instruction — see Table~\ref{tab:reasoning-variants}\,$\rangle$}\\
\mbox{}\\
{\color{tagcolor}<OUTPUT\_FORMAT>}\\
\ldots
\end{userprompt}
\caption{Position of the variant-specific instructions within the user prompt.}
\label{fig:prompt:variant_instructions}
\end{figure*}
 
We consider eight reasoning variants, described in Table~\ref{tab:reasoning-variants}. Each variant instantiates a well-established prompting strategy from the LLM-reasoning
literature, repurposed here as a source of \emph{controlled} prompt-level diversity for
the ensemble of Section~\ref{sec:llm-ensemble}; this design follows the diversity-of-thought paradigm~\citep{wangselfconsistency2023,naikDiversityOfThought2024}, in which majority-vote ensembles benefit from elicitations that follow heterogeneous cognitive paths rather than i.i.d.\ samples of the same path. Concretely, the variants span: structured criterion-by-criterion scoring with explicit intermediate steps (\textbf{Systematic}, after chain-of-thought prompting~\citep{weiChainOfThought2022}); user-centred simulation through persona instantiation (\textbf{Empathetic}, after in-context impersonation
 ~\citep{salewskiInContextImpersonation2023});
balanced two-sided enumeration (\textbf{Pros/Cons}) and steelman-style debate (\textbf{Adversarial}), both inspired by multi-agent debate
prompting~\citep{duImprovingMultiagentDebate2024,liangEncouragingDivergent2024} and here transposed to a single-LLM single-pass setting in which the model is asked to internalise competing perspectives within one elicitation; per-item decomposition into sub-judgments (\textbf{Item-level}, after problem-decomposition prompting~\citep{zhouLeastToMost2023,khotDecomposedPrompting2023}); critique-oriented identification of failure modes (\textbf{Fault-first}, after iterative self-critique paradigms~\citep{madaanSelfRefine2023}); holistic-then-analytic reasoning that first forms a global impression and then verifies against the rubric; and numerical rubric-based scoring in a form-filling paradigm (\textbf{Quantitative}, after G-Eval~\citep{liuGEval2023}).
The intent is that each strategy surfaces different aspects of list quality while grounding the final verdict in the same four-criterion rubric.

\begin{table*}[h!]
\centering
\small
\renewcommand{\arraystretch}{1.45}
\begin{tabularx}{\linewidth}{clX}
\toprule
\textbf{\#} & \textbf{Name} & \textbf{Reasoning-approach instruction} \\
\midrule
0 & Systematic
  & \textit{Using a structured, step-by-step approach, systematically evaluate
    each criterion before making your decision.} \\
1 & Empathetic
  & \textit{Put yourself in this user's shoes. Based on their interaction
    history, imagine what kind of experience they want next. Then evaluate
    which list would satisfy them most.} \\
2 & Pros/Cons
  & \textit{For each list, identify its top 3 strengths and top 3 weaknesses
    relative to this user's profile. Then weigh the pros against the cons to
    determine which list is superior.} \\
3 & Item-level
  & \textit{Analyse each individual item in both lists. For each item, assess
    how well it fits this user's profile. Then aggregate your item-level
    assessments to determine which list is stronger overall.} \\
4 & Fault-first
  & \textit{Start by identifying the most critical flaws in each list:
    fabricated items, major preference misalignments, redundancies. Then assess
    which list has fewer and less severe issues, complementing with positive
    qualities.} \\
5 & Impression
  & \textit{First, form an overall impression of each list as a complete
    recommendation package. Then verify your intuition by checking against the
    specific evaluation criteria. Confirm or revise your initial assessment.} \\
6 & Adversarial
  & \textit{First, construct the strongest possible argument that List~1 is the
    better choice. Then, construct the strongest possible argument that List~2
    is better. Finally, weigh both arguments to reach your verdict.} \\
7 & Quantitative
  & \textit{For each evaluation criterion, assign a score from 1 to 10 for
    each list. Compute the weighted sum and compare the totals. Let the numbers
    guide your verdict.} \\
\bottomrule
\end{tabularx}
\caption{The eight reasoning-approach instructions used across prompt variants.}
\label{tab:reasoning-variants}
\end{table*}

\section{Usage of Large Language Models (LLMs)}

We used LLMs for grammar, translation and spelling correction, as well as limited coding assistance. All generated code was manually reviewed and verified.

\end{document}